\pdfoutput=1
\documentclass[a4paper,USenglish,thm-restate,autoref]{lipics-v2021}

\hideLIPIcs
\nolinenumbers

\usepackage{import}

\usepackage{thmtools}
\usepackage{amsfonts}
\usepackage{amsmath}
\usepackage{mathtools}
\usepackage{amsthm}
\usepackage{amssymb} %

\usepackage{microtype}

\usepackage{footnote}
\makesavenoteenv{tabular}
\makesavenoteenv{table}
\usepackage{tablefootnote}

\usepackage[dvipsnames]{xcolor}
\usepackage{graphicx}

\usepackage{subcaption}
\usepackage{float}
\definecolor{linkblue}{HTML}{001487}

\usepackage{tikz}
\usepackage{shortsym}

\usepackage{pifont}%
\newcommand{\cmark}{\ding{51}}%
\newcommand{\xmark}{\ding{55}}%

\usetikzlibrary{arrows.meta}
\usetikzlibrary{decorations.pathreplacing} %

\usepackage{xspace}
\usepackage{csquotes}
\usepackage{subcaption}

\usepackage{booktabs}
\usepackage{nicefrac}

\usepackage[
    lambda,
	operators,
	advantage,
	sets,
	adversary,
	landau,
	probability,
	notions,	
	logic,
	ff,
	mm,
	primitives,
	events,
	complexity,
	asymptotics,
	keys
    ]{cryptocode}

\newcommand{\seed}{\mathsf{seed}}
\newcommand{\epoch}{\mathsf{epoch}}

\newcommand{\lock}{\mathsf{lock}}
\newcommand{\candidate}{\mathsf{candidate}}
\newcommand{\GPE}[1]{\mathsf{GPE_{#1}}}
\newcommand{\Myadversary}{{External} }
\newcommand{\myadversary}{{external} }
\newcommand{\exehorizon}{\mathrm{T_{horizon}}}
\newcommand{\bE}{{\bf E}}

\newcommand{\Tf}{\ensuremath{T_{\mathrm{f}}}}
\newcommand{\Tb}{\ensuremath{T_{\mathrm{b}}}}

\theoremstyle{plain}
\newtheorem*{theorem*}{Theorem}
\newtheorem*{lemma*}{Lemma}

\newcommand{\negli}[1]{\mathsf{negl}(#1)}

\newcommand{\val}{\mathsf{val}}
\newcommand{\Tr}{\mathrm{Tr}}
\def\rmV{\mathrm{V}}
\def\rmS{\mathrm{S}}
\def\rmD{\mathrm{D}}
\newcommand{\reg}[1]{\mathsf{#1}}
\newcommand{\view}{\mathsf{view}}

\newcommand{\ip}[1]{\langle #1 \rangle}
\newcommand{\cA}{\ensuremath{\mathcal{A}}}

\newcommand{\cF}{\ensuremath{\mathcal{F}}}

\newcommand{\cL}{\ensuremath{\mathcal{L}}}

\newcommand{\cO}{\ensuremath{\mathcal{O}}}
\newcommand{\cP}{\ensuremath{\mathcal{P}}}

\makeatletter
\AddToHook{env/algorithmic/begin}{\def\@currentcounter{ALG@line}}
\makeatother

\usepackage[sort&compress,capitalize,nameinlink]{cleveref}

\AtBeginEnvironment{appendices}{%
    \crefalias{section}{appendix}%
    \crefalias{subsection}{subappendix}%
    \crefalias{subsubsection}{subsubappendix}%
    \crefalias{subsubsubsection}{subsubsubappendix}%
}
\AddToHook{cmd/appendix/after}{%
    \crefalias{section}{appendix}%
    \crefalias{subsection}{subappendix}%
    \crefalias{subsubsection}{subsubappendix}%
    \crefalias{subsubsubsection}{subsubsubappendix}%
}

\hypersetup{hypertexnames=false}

\crefalias{ALG@line}{line}

\crefname{figure}{Fig.}{Figs.}
\Crefname{figure}{Fig.}{Figs.}

\crefname{table}{Tab.}{Tabs.}
\Crefname{table}{Tab.}{Tabs.}

\crefname{section}{Sec.}{Secs.}
\Crefname{section}{Sec.}{Secs.}
\crefname{subsection}{Sec.}{Secs.}
\Crefname{subsection}{Sec.}{Secs.}
\crefname{subsubsection}{Sec.}{Secs.}
\Crefname{subsubsection}{Sec.}{Secs.}
\crefname{subsubsubsection}{Sec.}{Secs.}
\Crefname{subsubsubsection}{Sec.}{Secs.}
\crefname{appendix}{App.}{Apps.}
\Crefname{appendix}{App.}{Apps.}
\crefname{subappendix}{App.}{Apps.}
\Crefname{subappendix}{App.}{Apps.}
\crefname{subsubappendix}{App.}{Apps.}
\Crefname{subsubappendix}{App.}{Apps.}
\crefname{subsubsubappendix}{App.}{Apps.}
\Crefname{subsubsubappendix}{App.}{Apps.}

\crefname{algorithm}{Alg.}{Algs.}
\Crefname{algorithm}{Alg.}{Algs.}
\crefname{line}{ln.}{lns.}
\Crefname{line}{ln.}{lns.}

\crefname{proposition}{Prop.}{Props.}
\Crefname{proposition}{Prop.}{Props.}
\crefname{lemma}{Lem.}{Lems.}
\Crefname{lemma}{Lem.}{Lems.}
\crefname{theorem}{Thm.}{Thms.}
\Crefname{theorem}{Thm.}{Thms.}
\crefname{corollary}{Cor.}{Cors.}
\Crefname{corollary}{Cor.}{Cors.}
\crefname{definition}{Def.}{Defs.}
\Crefname{definition}{Def.}{Defs.}

\crefname{claim}{Clm.}{Clms.}
\Crefname{claim}{Clm.}{Clms.}

\usepackage{algorithm}
\usepackage{algorithmicx}
\usepackage[noend]{algpseudocode}

\makeatletter
\AddToHook{env/algorithmic/begin}{\def\@currentcounter{ALG@line}}
\makeatother

\algnewcommand{\LineComment}[1]{\State {\textcolor{gray}{/\!/ #1}}}

\algrenewcommand{\alglinenumber}[1]{\scriptsize\textcolor{gray}{\texttt{#1}}}
\algrenewcommand{\algorithmicindent}{1em}

\algnewcommand{\algfontsize}[0]{}
\AddToHook{env/algorithmic/begin}{\algfontsize}

\algnewcommand{\algorithmicswitch}{\textbf{switch}}
\algdef{SE}[SWITCH]{Switch}{EndSwitch}[1]{\algorithmicswitch\ #1\ \algorithmicdo}{\algorithmicend\ \algorithmicswitch}%
\algtext*{EndSwitch}%

\algnewcommand{\algorithmiccase}{\textbf{case}}
\algdef{SE}[CASE]{Case}{EndCase}[1]{\algorithmiccase\ #1}{\algorithmicend\ \algorithmiccase}%
\algtext*{EndCase}%

\algnewcommand{\algorithmicon}{\textbf{on}}
\algdef{SE}[ON]{On}{EndOn}[1]{\algorithmicon\ #1\ \algorithmicdo}{\algorithmicend\ \algorithmicon}%
\algtext*{EndOn}%

\algnewcommand{\algorithmicupon}{\textbf{upon}}
\algdef{SE}[UPON]{Upon}{EndUpon}[1]{\algorithmicupon\ #1\ \algorithmicdo}{\algorithmicend\ \algorithmicupon}%
\algtext*{EndUpon}%

\algnewcommand{\algorithmicat}{\textbf{at}}
\algdef{SE}[AT]{At}{EndAt}[1]{\algorithmicat\ #1\ \algorithmicdo}{\algorithmicend\ \algorithmicat}%
\algtext*{EndAt}%

\algnewcommand{\algorithmicrealfunction}{\textbf{function}}
\algdef{SE}[REALFUNCTION]{RealFunction}{EndRealFunction}[1]{\algorithmicrealfunction\ #1\ \algorithmicdo}{\algorithmicend\ \algorithmicrealfunction}%
\algtext*{EndRealFunction}%

\algnewcommand{\algorithmicthroughout}{\textbf{do throughout}}
\algdef{SE}[Throughout]{Throughout}{EndThroughout}[1]{\algorithmicthroughout\ #1\ \algorithmicdo}{\algorithmicend\ \algorithmicthroughout}%
\algtext*{EndThroughout}%

\algnewcommand{\algorithmictry}{\textbf{try}}
\algnewcommand{\algorithmiccatch}{\textbf{catch}}
\algdef{SE}[TRY]{Try}{EndTry}{\algorithmictry\ \algorithmicdo}{\algorithmicend\ \algorithmictry}
\algdef{C}[TRY]{TRY}{Catch}[1]{\algorithmiccatch\ #1\ \algorithmicdo}
\algtext*{EndTry}

\algrenewcommand{\algorithmicdo}{}
\algrenewcommand{\algorithmicthen}{}

\algnewcommand{\algorithmicgoto}{\textbf{goto}}%
\algnewcommand{\Goto}[1]{\algorithmicgoto~\ref{#1}}%

\algnewcommand{\algorithmicassert}{\textbf{assert}}%
\algnewcommand{\Assert}[1]{\algorithmicassert~{#1}}%

\algnewcommand{\algorithmicbreak}{\textbf{break}}%
\algnewcommand{\Break}[0]{\algorithmicbreak}%
\algnewcommand{\BreakOutOf}[1]{\algorithmicbreak~out~of~#1}%

\algnewcommand{\algorithmicwaiton}{\textbf{wait on}}%
\algnewcommand{\WaitOn}[1]{\algorithmicwaiton~{#1}}%

\algnewcommand{\InlineRequire}[1]{\textbf{require} {#1}}

\algblock{ManualIndent}{EndManualIndent}
\algnotext{ManualIndent}
\algnotext{EndManualIndent}

\algdef{SE}[GENERICBLOCK]{GenericBlock}{EndGenericBlock}[1]{#1}{}%
\algtext*{EndGenericBlock}%

\usetikzlibrary{patterns}

\tikzset{blockchain/.style={
        x=0.5cm,
        y=0.55cm,
        node distance=0.5cm,
        block/.style = {
                minimum width=0.3cm,
                minimum height=0.3cm,
                draw,
                shade,
                top color=white,
                bottom color=black!10,
                inner sep=0,
            },
        block-adv/.style = {
                block,
                bottom color=Red!50,
                draw=Red!50!black,
            },
        block-hon/.style = {
                block,
                bottom color=Green!50!black!50,
                draw=Green!50!black,
            },
        block-blank/.style = {
                minimum width=0.3cm,
                minimum height=0.3cm,
                rounded corners,
                inner sep=0,
            },
        link/.style = {
                -latex,
            },
        link-adv/.style = {
                link,
            },
        link-hon/.style = {
                link,
            },
    }
}

\definecolor{myParula01Blue}{RGB}{0,114,189}
\definecolor{myParula02Orange}{RGB}{217,83,25}
\definecolor{myParula03Yellow}{RGB}{237,177,32}
\definecolor{myParula04Purple}{RGB}{126,47,142}
\definecolor{myParula05Green}{RGB}{119,172,48}
\definecolor{myParula06LightBlue}{RGB}{77,190,238}
\definecolor{myParula07Red}{RGB}{162,20,47}

\title{Fully-Fluctuating Participation in Sleepy Consensus}
\author{Yuval Efron}{Columbia University, New York, NY, USA}{ye2210@columbia.edu}{https://orcid.org/0000-0003-0882-9342}{}
\author{Joachim Neu}{a16z Crypto Research, New York, NY, USA}{jneu@a16z.com}{https://orcid.org/0000-0002-9777-6168}{}
\author{Toniann Pitassi}{Columbia University, New York, NY, USA}{tp2684@columbia.edu}{}{}
\authorrunning{Y.\ Efron, J.\ Neu, and T.\ Pitassi}

\acknowledgements{We thank 
Joseph Bonneau,
Javier Nieto,
and
Ling Ren
for fruitful discussions.}


\Copyright{Yuval Efron, Joachim Neu, and Toniann Pitassi}

\EventEditors{Zeta Avarikioti and Nicolas Christin}
\EventNoEds{2}
\EventLongTitle{7th Conference on Advances in Financial Technologies (AFT 2025)}
\EventShortTitle{AFT 2025}
\EventAcronym{AFT}
\EventYear{2025}
\EventDate{October 8--10, 2025}
\EventLocation{Pittsburgh, PA, USA}
\EventLogo{}
\SeriesVolume{354}
\ArticleNo{7}

\begin{document}
\maketitle
\begin{abstract}
Proof-of-work allows Bitcoin 
to boast security amidst arbitrary fluctuations in participation of miners throughout time, so long as, at any point in time, a majority of hash power is honest.
In recent years, however, the pendulum has shifted in favor of proof-of-stake-based consensus protocols.
There, the sleepy model
is the most prominent 
model for handling fluctuating participation of nodes.
However, to date, no protocol in the sleepy model rivals Bitcoin 
in its robustness to drastic fluctuations in participation levels,
with state-of-the-art protocols making various restrictive assumptions.
In this work, we present a new adversary model, called \emph{external adversary}. Intuitively, in our model, corrupt nodes do not divulge information about their secret keys.
In this model, we show that
protocols in the sleepy model can meaningfully claim to remain secure
against fully fluctuating participation, without compromising efficiency or corruption resilience. 
Our adversary model is quite natural, and arguably naturally captures the process via which malicious behavior arises in protocols, as opposed to traditional worst-case modeling. 
On top of which, the model is also theoretically appealing, circumventing a barrier established in a recent work of 
Malkhi, Momose, and Ren.

\end{abstract}

\section{Introduction}
\label{sec:intro}

Byzantine-fault tolerant (BFT) consensus, the problem of reaching agreement on a value amongst $n$ nodes, of which a fraction are corrupted and controlled by an \emph{adversary}, is one of the most fundamental problems in distributed computing, with research spanning over four decades~\cite{LamportSP82,PeaseSL80}. The last decade has seen a significant growth in work and interest in the problem, mainly motivated by the pivotal role that consensus protocols play in the design and analysis of blockchains~\cite{Bitcoin,Gramoli20,CrainGLR18,AbrahamMN0S17,Thunderella,FitziGKR18}. 

Traditionally, it is assumed that apart from the corrupt nodes, all other nodes follow the protocol at all times. This assumption, while justified for applications such as data centers, is arguably optimistic in the context of blockchains, where the participants in the protocol may come and go at arbitrary points in time.  Specifically, the desiderata for many blockchain protocols, such as Bitcoin and Ethereum, requires maintaining security even in the presence of \emph{many unexpected temporary crash faults}, in which some subset of the nodes may go offline for an undetermined interval of time. 
In particular, the protocol may experience alternating periods of high participation and low participation.\footnote{Throughout, we assume a fixed set of \emph{potential} participants \emph{known} to the protocol; i.e.,\ the set of nodes allowed to partake 
is fixed, and fluctuation in participation occurs w.r.t.\ that set. \emph{Reconfiguration} of the nodes partaking in the protocol is orthogonal to the 
\emph{unexpected} and \emph{wide} fluctuations in participation considered here, and reconfiguration is expressly \emph{not} considered in this work.} 
The seminal Bitcoin protocol~\cite{Bitcoin}, also known as proof-of-work Nakamoto consensus, 
was the first protocol to achieve 
secure consensus under such circumstances,
allowing participants in the protocol to come and go as they please so long as an honest majority (of hash-based mining power)
is maintained at any point in time. 
Its deftly formalized proof-of-work framework allowed for protocol design which is completely agnostic to the total number of participating nodes (corrupt or honest!) \emph{across} present, past, or future. 
For example, if tomorrow the honest participation in Bitcoin drops to a level below the corrupt participation \emph{today}, it would not render Bitcoin insecure (so long as corrupt participation decreases accordingly). 
The same holds for if corrupt participation \emph{tomorrow} exceeds honest participation \emph{today} (assuming honest participation increases accordingly).

As the years went on, 
proof-of-work
ceded its spot in the limelight to proof-of-stake, which presently is the dominant approach to Sybil-resistance among 
blockchain protocols. 
In the translation, however, resilience to fluctuation in participation was lost. In the context of proof-of-stake, a formal model that captures fluctuating participation is the \emph{sleepy} consensus model, introduced in \cite{PassS17} by Pass and Shi. This model is the focus of our work. In the sleepy model, there is a fixed set of $n$ nodes, identified via a public key infrastructure (PKI). Besides being classified as \emph{honest} or \emph{corrupt}, nodes can alternate (adversarially) between two states at each point in time: 
\emph{awake}, or 
\emph{asleep}. 
Awake nodes can participate in the protocol normally, i.e., perform local computation, send and receive messages. On the other hand, asleep nodes cannot perform any computation and cannot engage in the protocol in any way, until they wake up. While a proof-of-stake instantiation of Nakamoto's protocol 
was proven secure~\cite{PassS17} in the sleepy model, the proof assumes significant restrictions on fluctuating participation. Namely, that honest participation at \emph{any} time exceeds corrupt participation at \emph{all} times. Such a concession perhaps felt inherent, with many follow-up works in the sleepy model making the same assumption~\cite{PassS17,GoyalLR21,DACS,Ouroboros,Thunderella,Snowwhite,FitziGKR18}.
This intuition for the necessity of such an assumption was formalized in~\cite{MalkhiM023}, which showed that indeed any consensus protocol that does not employ time-based cryptography (e.g., verifiable delay functions~\cite{boneh2018verifiable}, henceforth VDF) must heavily restrict fluctuating participation.

The impossibility~\cite{MalkhiM023} is established by the following \emph{key transfer} attack.
Consider a consensus protocol that makes use of a PKI setup, where it is assumed that prior to the initiation of the protocol, all nodes were privately handed secret-key/public-key pairs. Pondering on the PKI assumption for a moment, however, one observes that the notion of an \emph{asleep corrupt node} makes little sense, as any corrupt node that is awake for even a single point in time, and whose behaviour is completely controlled by the adversary, can simply broadcast its secret key to all other corrupt nodes, and then be put to sleep. Thus allowing essentially a \emph{single} awake corrupt node to simulate a behaviour indistinguishable from all corrupt nodes being awake. Such an attack clearly renders a protocol vulnerable to the case where honest participation decreases over time.

What remains is perhaps an unfortunate state of affairs, in which proof-of-stake protocols seem to inherently carry the burden of rigid participation constraints. The goal of this work is to challenge this status quo. We do so by revisiting the adversary model. Specifically,~\cite{MalkhiM023} has already observed that such a strong adversary model capable of executing the key transfer attack may be a tad unrealistic:
``The (standard) adversary assumption, where nodes can extract all corrupt nodes' private states, is too strong in practice.
In particular in the proof-of-stake protocols, it is highly unlikely that corrupt nodes hand off their secret keys to the adversary (or other corrupt nodes) at the risk of losing their entire stake.''~\cite{MalkhiM023}
They conclude by raising the problem of achieving sleepy consensus under fully-fluctuating corrupt nodes in a ``more realistic adversarial model.''

Motivated by similar principles, we consider in this work a mildly relaxed adversary model, which we refer to as an \emph{\myadversary adversary}. Informally, an external adversary is external to the network and the protocol; the adversary can observe the protocol from a birds eye view. In turn, corrupt nodes are not inherently malicious, but opportunistic: they are recruited by the external adversary to perform arbitrary behavior in the protocol, but won't divulge information about their own secret key to any other node. We believe it is much more likely that a malicious entity would be able to ``rent'' keys in the protocol, by means of recruiting nodes to engage in Byzantine behavior, rather than convince nodes to give up their secret keys. Stating a formal model that accurately captures the above intuition turns out to be quite simple, but does requires some care, and we tend to it in \cref{sec:model_and_defs}. Not only does this model still capture a rich plethora of attack vectors pertinent in practice, but it is also theoretically appealing: Disallowing corrupt nodes to share information about their secret keys with one another is precisely the minimal relaxation required to render the key transfer attack infeasible.

\begin{figure}
    \centering
    \begin{subfigure}[b]{0.49\linewidth}
        \centering
        \begin{tikzpicture}[x=4.5cm,y=3.5cm]
            \footnotesize
            
            \pgfdeclarelayer{axis}
            \pgfdeclarelayer{plots}
            \pgfsetlayers{plots,axis}
    
            \begin{pgfonlayer}{axis}
                \draw [-Latex] (0,0) -- (1.05,0) node [midway,below] {Time};
                \draw [-Latex] (0,0) -- (0,0.95) node [midway,above,rotate=90] {Participation};
            \end{pgfonlayer}

            \begin{pgfonlayer}{plots}
                    \draw [Green,thick] plot [smooth,domain=0:1,samples=100] (\x,{0.5 + (-3*(\x-0.5)^2+1)*0.2*sin(1440*\x)}) node [right] {$h(t)$};
                    \draw [Red,thick] plot [smooth,domain=0:1,samples=100] (\x,{0.25}) node [right] {$c(t)$};
            \end{pgfonlayer}
        \end{tikzpicture}
        \caption{Consistent}
        \label{fig:sleepy_settings_consistent}
    \end{subfigure}%
    \hfil
    \begin{subfigure}[b]{0.49\linewidth}
        \centering
        \begin{tikzpicture}[x=4.5cm,y=3.5cm]
            \footnotesize
            
            \pgfdeclarelayer{axis}
            \pgfdeclarelayer{plots}
            \pgfsetlayers{plots,axis}
    
            \begin{pgfonlayer}{axis}
                \draw [-Latex] (0,0) -- (1.05,0) node [midway,below] {Time};
                \draw [-Latex] (0,0) -- (0,0.95) node [midway,above,rotate=90] {Participation};
            \end{pgfonlayer}

            \begin{pgfonlayer}{plots}
                    \draw [Green,thick] plot [smooth,domain=0:1,samples=100] (\x,{0.6 + (-3*(\x-0.5)^2+1)*0.2*sin(1440*\x)}) node [right] {$h(t)$};
                    \draw [Red,thick] plot [smooth,domain=0:1,samples=100] (\x,{0.4 + (-3*(\x-0.5)^2+1)*0.2*sin(1440*\x)}) node [right] {$c(t)$};
            \end{pgfonlayer}
        \end{tikzpicture}
        \caption{Fully-fluctuating}
        \label{fig:sleepy_settings_fullyfluctuating}
    \end{subfigure}%
    \\[1em]
    \begin{subfigure}[b]{0.49\linewidth}
        \centering
        \begin{tikzpicture}[x=4.5cm,y=3.5cm]
            \footnotesize
            
            \pgfdeclarelayer{axis}
            \pgfdeclarelayer{plots}
            \pgfsetlayers{plots,axis}
    
            \begin{pgfonlayer}{axis}
                \draw [-Latex] (0,0) -- (1.05,0) node [midway,below] {Time};
                \draw [-Latex] (0,0) -- (0,0.95) node [midway,above,rotate=90] {Participation};
            \end{pgfonlayer}

            \begin{pgfonlayer}{plots}
                    \draw [Green,thick] plot [smooth,domain=0:1,samples=100] (\x,{0.5*(\x-1)^2+0.3 + (-3*(\x-0.5)^2+1)*0.07*sin(1440*\x)}) node [right] {$h(t)$};
                    \draw [Red,thick] plot [smooth,domain=0:1,samples=100] (\x,{0.5*(\x-1)^2+0.1 + (-3*(\x-0.5)^2+1)*0.07*sin(1440*\x)}) node [right] {$c(t)$};
                    \draw [dashed] plot [smooth,domain=0:1,samples=100] (\x,{0.5*(\x-1)^2+0.2}) node [right] {$x(t)$};
            \end{pgfonlayer}
        \end{tikzpicture}
        \caption{Decreasing}
        \label{fig:sleepy_settings_decreasing}
    \end{subfigure}%
    \hfil
    \begin{subfigure}[b]{0.49\linewidth}
        \centering
        \begin{tikzpicture}[x=4.5cm,y=3.5cm]
            \footnotesize
            
            \pgfdeclarelayer{axis}
            \pgfdeclarelayer{plots}
            \pgfsetlayers{plots,axis}
    
            \begin{pgfonlayer}{axis}
                \draw [-Latex] (0,0) -- (1.05,0) node [midway,below] {Time};
                \draw [-Latex] (0,0) -- (0,0.95) node [midway,above,rotate=90] {Participation};
            \end{pgfonlayer}

            \begin{pgfonlayer}{plots}
                    \draw [Green,thick] plot [smooth,domain=0:1,samples=100] (\x,{0.5*((1-\x)-1)^2+0.3 + (-3*((1-\x)-0.5)^2+1)*0.07*sin(1440*(1-\x))}) node [right] {$h(t)$};
                    \draw [Red,thick] plot [smooth,domain=0:1,samples=100] (\x,{0.5*((1-\x)-1)^2+0.1 + (-3*((1-\x)-0.5)^2+1)*0.07*sin(1440*(1-\x))}) node [right] {$c(t)$};
                    \draw [dashed] plot [smooth,domain=0:1,samples=100] (\x,{0.5*((1-\x)-1)^2+0.2}) node [right] {$x(t)$};
            \end{pgfonlayer}
        \end{tikzpicture}
        \caption{Increasing}
        \label{fig:sleepy_settings_increasing}
    \end{subfigure}%
    \caption{Examples of possible participation rates for each of the discussed settings. The graphs depict the number of participating nodes at any point in time, with green curves depicting honest participation, and red curves depicting corrupt participation. Note that in all four settings, honest participation exceeds corrupt participation at any given point in time. On top of that, in the consistent case, note that \emph{any} point on the honest curve dominates the corrupt curve at \emph{all} points in time. Denoting by $h(t)$ and $c(t)$ the honest and corrupt participation at time $t$, the increasing (decreasing) setting can be characterized by the existence of a monotone increasing (decreasing) function $x(t)$ such that $c(t)<x(t)<h(t)$ holds at all times. 
    The fully-fluctuating case poses no additional restrictions.}
    \label{fig:sleepy_settings}
\end{figure}
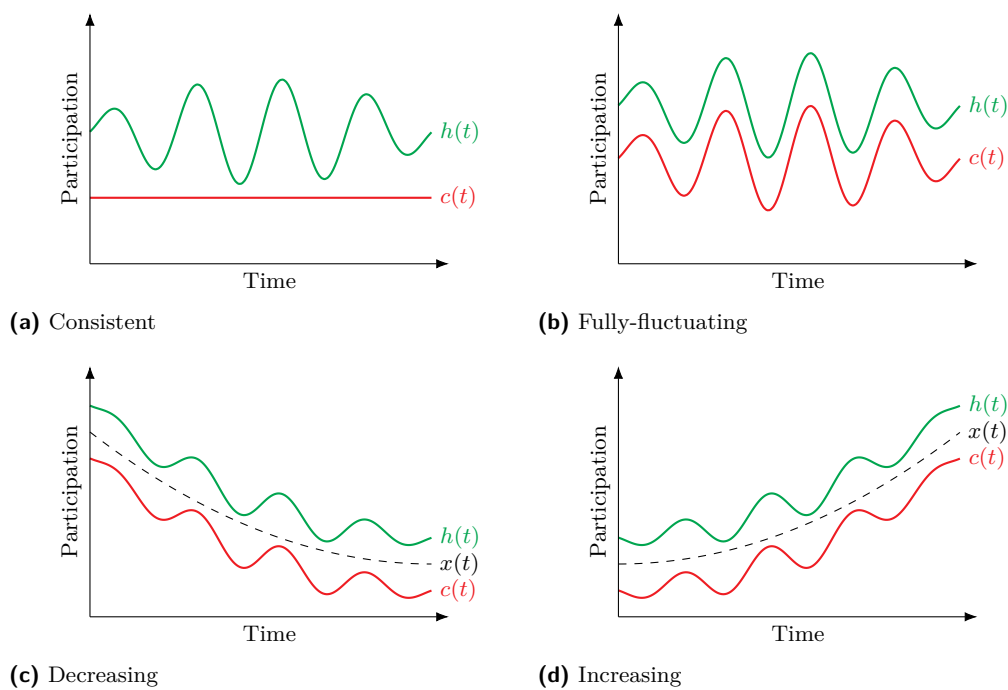

\subparagraph{The settings.} 
Pondering on the notion of fluctuating participation of nodes in the protocol, one can identify several interesting settings to consider. See also \cref{fig:sleepy_settings}.
\begin{enumerate}
    \item \textbf{Consistent participation.} 
    (\cref{fig:sleepy_settings_consistent})
    This is the most common assumption in the context of the sleepy model~\cite{PassS17,GoyalLR21,DACS,Ouroboros,Thunderella,Snowwhite,FitziGKR18}, where it is assumed that new (i.e., after time 0) corrupt nodes may never wake up, and corrupt nodes may never go to sleep. Additionally, an honest majority is assumed to hold at all times. In other words, for \emph{any} point time, honest participation exceeds corrupt participation at \emph{all} times.
    \item \textbf{Increasing participation.} 
    (\cref{fig:sleepy_settings_increasing})
    In this setting, participation of corrupt nodes is guaranteed to be non decreasing as time progresses. In other words, new corrupt nodes may wake up throughout the protocol, but may no go to sleep once woken up. Note that in this setting, even assuming an honest majority at any point in time, it could be that as time goes on, corrupt participation exceeds honest participation in \emph{earlier} points in time. As a matter of fact, this setting has received a fair amount of attention recently~\cite{Momose022,MalkhiM023,Goldfish,StreamlineSleepy,Posat}, in several works showcasing efficient consensus protocols for this setting.
    \item \textbf{Decaying participation.} 
    (\cref{fig:sleepy_settings_decreasing})
    In this setting, participation of corrupt nodes is guaranteed to be non increasing as time progresses. In other words, corrupt nodes may go to sleep throughout the protocol, but can never be newly woken up. In this setting, even with an honest majority at any point in time, it could be that as time goes on, corrupt participation in the \emph{past} exceeds \emph{current} honest participation.
    \item \textbf{Fully-fluctuating.} 
    (\cref{fig:sleepy_settings_fullyfluctuating})
    This is the most general setting in the context of the sleepy model. In this setting, participation of both honest and corrupt nodes can both increase and decrease over time, to the discretion of the adversary. In particular, any of the scenarios mentioned in the previous settings can also occur in this setting. 
\end{enumerate}

\subsection{Main results}
\label{ssec:contributions}

Our results establish that assuming an \myadversary adversary allows protocols in the sleepy model to be endowed with resilience to fluctuating participation of nodes, rivaling Nakamoto in its graceful handling of unknown participation patterns.

\subparagraph{(1) Sleepy consensus under decaying participation.}
Our first result establishes the utility of the \myadversary adversary model by showcasing a separation from the standard adversary model. Namely, consider the case where participation of both honest and corrupt nodes decreases over time. In particular, for any point in time, \emph{future} honest participation may be  eclipsed by \emph{current} corrupt participation. 
Recent work~\cite{MalkhiM023} showed that in this setting, no protocol can be secure in the presence of an adversary that can carry out key transfer.

Our first result shows that against an \myadversary adversary, a consensus protocol resilient to decaying participation is not only feasible, but can be made highly efficient and requires only standard cryptographic assumptions.

\begin{theorem*}[Informal, see \cref{thrm:decaying_participation_protocol}]\label{thrm:decaying_participation_informal}
    Assuming a PKI, a VRF, and an \myadversary adversary, we present a randomized protocol 
    solving consensus
    in the decaying participation setting with expected constant latency. 
\end{theorem*}

\subparagraph{(2) Consensus under fully-fluctuating participation.} 
\Cref{thrm:decaying_participation_protocol} establishes the strength of the \myadversary adversary model from the lens of protocol design, allowing us to obtain provable security guarantees known to be impossible under the standard notion of an adversary, that is able to carry out the key transfer attack.
We now turn our attention to the more general setting of \emph{fully-fluctuating} participation, in which participation of honest and corrupt nodes can both increase and decrease over time, to the discretion of the adversary. Designing a protocol secure in the fully-fluctuating setting in a relaxed adversary model was explicitly stated as an open problem in~\cite{MalkhiM023}. We show that with the help of VDFs, one can design an efficient consensus protocol for the fully-fluctuating participation setting, secure against an \myadversary adversary.

\begin{theorem*}[Informal, see \cref{thrm:main_thrm}]\label{thrm:non_collude_adversary_informal}
    Assuming a PKI, a VRF, a VDF, and an \myadversary adversary, we present a randomized protocol
    solving consensus that tolerates fully-fluctuating participation with expected constant latency.
\end{theorem*}

\Cref{table:protocols} summarizes our contributions and gives full context of previous work using the formal notation we introduce in \cref{sec:model_and_defs}. See \cref{ssec:related_work} for an in-depth overview of related work.

\subsection{Key techniques}
\label{ssec:key_techniques}

\subparagraph{The challenge.}  
Even in non-sleepy synchrony settings~\cite{adversarymajorityclients}, any consensus protocol must assume that the adversary is limited in corrupting only a minority of the participants. The sleepy model, being a generalization of the traditional setting, clearly must also abide by an assumption of similar flavour. Note further that in the sleepy model, the awake/asleep status of every node at every point in time is also dictated by the adversary. 
Thus the changing nature of node participation in the sleepy model makes such an assumption non-trivial to state. Arguably the minimal assumption one could make, is that at each point in time, honest participation exceeds corrupt participation. E.g., this is the assumption for which Nakamoto consensus~\cite{Bitcoin} is proven secure in the proof-of-work setting. In the translation to proof-of-stake, however, such a minimal assumption forces any purported protocol to be resilient against the following two attack vectors. We emphasize that these attacks remain a concern in the external adversary model, and are described to go through within the confines of the model. To envisage the attack vectors in a tangible manner, we consider here as a running example the proof-of-stake version of Nakamoto consensus.

\begin{figure}[tb]
    \centering
    \begin{subfigure}[b]{\linewidth}
        \centering
        \begin{tikzpicture}[x=2.5cm,y=2cm]
            \scriptsize

            \begin{scope}
                \draw [Green,thick] plot [smooth,domain=0:1,samples=100] (\x,{0.5*(\x)^2+0.3 + (-3*(\x-0.5)^2+1)*0.07*sin(1440*\x)});
                \draw [Red,thick] plot [smooth,domain=0:1,samples=100] (\x,{0.5*(\x)^2+0.1 + (-3*(\x-0.5)^2+1)*0.07*sin(1440*\x)});
                \draw [densely dotted] (0,0.6) -- (1,0.6) node [pos=0,left] {$c_0$};

                \draw [-latex] (0,0) -- (1.05,0) node [midway,below] {Time};
                \draw [-latex] (0,0) -- (0,0.95) node [midway,above,xshift=-1.5em,rotate=90] {Participation};

                \node [yshift=1cm] at (0.5,1) {\textsc{True participation}};
            \end{scope}

            \begin{scope}[xshift=4.5cm]
                \draw [fill=Green,fill opacity=0.2,draw=none] plot [smooth,domain=0:1,samples=100] (\x,{0.5*(\x)^2+0.3 + (-3*(\x-0.5)^2+1)*0.07*sin(1440*\x)}) -- (1,0) -- (0,0) -- cycle;
                \draw [fill=Red,fill opacity=0.2,draw=none] plot [smooth,domain=0:1,samples=100] (\x,{0.6}) -- (1,0) -- (0,0) -- cycle;
                \draw [Green,thick] plot [smooth,domain=0:1,samples=100] (\x,{0.5*(\x)^2+0.3 + (-3*(\x-0.5)^2+1)*0.07*sin(1440*\x)});
                \draw [Red,thick] plot [smooth,domain=0:1,samples=100] (\x,{0.6});
                \draw [densely dotted,draw=none] (0,0.6) -- (1,0.6) node [pos=0,left] {$c_0$};

                \draw [-latex] (0,0) -- (1.05,0) node [midway,below] {Time};
                \draw [-latex] (0,0) -- (0,0.95) node [midway,above,xshift=-1.5em,rotate=90] {Participation};

                \node [yshift=1em] at (0.5,1) {\tikz[x=1em,y=1em,baseline=0.2em]{ \draw [fill=Green,fill opacity=0.2,Green,thick] (0,0) rectangle (1,1); } $= \int h < \int c =$ \tikz[x=1em,y=1em,baseline=0.2em]{ \draw [fill=Red,fill opacity=0.2,Red,thick] (0,0) rectangle (1,1); }};

                \node [yshift=1cm] at (0.5,1) {\textsc{Perceived participation}};
            \end{scope}

            \begin{scope}[xshift=10cm]
                \begin{scope}[yshift=2.2cm,blockchain,x=2em,y=-2em]
                    \node [block,minimum size=1.2em] (G) at (0,0) {$B_0$};
                    \foreach \i in {1,...,3} {
                        \node [block-hon] (H\i) at (-1,\i) {};
                    }
                    \foreach \i in {2,3} {
                        \pgfmathtruncatemacro{\prev}{\i-1}
                        \draw [link] (H\i) -- (H\prev);
                    }
                    \draw [link] (H1) -- (G);
                    \foreach \i in {1,...,4} {
                        \node [block-adv] (A\i) at (1,\i) {};
                    }
                    \foreach \i in {2,3,4} {
                        \pgfmathtruncatemacro{\prev}{\i-1}
                        \draw [link] (A\i) -- (A\prev);
                    }
                    \draw [link] (A1) -- (G);
                    \draw [|<->|] ([xshift=-1em]H1.north west) -- ([xshift=-1em]H3.south west) node [midway,left] {$\int h$};
                    \draw [|<->|] ([xshift=1em]A1.north east) -- ([xshift=1em]A4.south east) node [midway,right] {$\int c$};
                \end{scope}

                \node [yshift=1cm] at (0,1) {\textsc{Blockchain}};
            \end{scope}

        \end{tikzpicture}
        \caption{Backward simulation attack}
        \label{fig:backward_simulation}
    \end{subfigure}
    \\[1em]
    \begin{subfigure}[b]{\linewidth}
        \centering
        \begin{tikzpicture}[x=2.5cm,y=2cm]
            \scriptsize

            \begin{scope}
                \draw [Green,thick] plot [smooth,domain=0:1,samples=100] (\x,{0.5*(\x-1)^2+0.3 + (-3*(\x-0.5)^2+1)*0.07*sin(1440*\x)});
                \draw [Red,thick] plot [smooth,domain=0:1,samples=100] (\x,{0.5*(\x-1)^2+0.1 + (-3*(\x-0.5)^2+1)*0.07*sin(1440*\x)});
                \draw [densely dotted] (0,0.6) -- (1,0.6) node [pos=0,left] {$c_0$};

                \draw [-latex] (0,0) -- (1.05,0) node [midway,below] {Time};
                \draw [-latex] (0,0) -- (0,0.95) node [midway,above,xshift=-1.5em,rotate=90] {Participation};

                \node [yshift=1cm] at (0.5,1) {\textsc{True participation}};
            \end{scope}

            \begin{scope}[xshift=4.5cm]
                \draw [fill=Green,fill opacity=0.2,draw=none] plot [smooth,domain=0:1,samples=100] (\x,{0.5*(\x-1)^2+0.3 + (-3*(\x-0.5)^2+1)*0.07*sin(1440*\x)}) -- (1,0) -- (0,0) -- cycle;
                \draw [fill=Red,fill opacity=0.2,draw=none] plot [smooth,domain=0:1,samples=100] (\x,{0.6}) -- (1,0) -- (0,0) -- cycle;
                \draw [Green,thick] plot [smooth,domain=0:1,samples=100] (\x,{0.5*(\x-1)^2+0.3 + (-3*(\x-0.5)^2+1)*0.07*sin(1440*\x)});
                \draw [Red,thick] plot [smooth,domain=0:1,samples=100] (\x,{0.6});
                \draw [densely dotted] (0,0.6) -- (1,0.6) node [pos=0,left] {$c_0$};

                \draw [-latex] (0,0) -- (1.05,0) node [midway,below] {Time};
                \draw [-latex] (0,0) -- (0,0.95) node [midway,above,xshift=-1.5em,rotate=90] {Participation};

                \node [yshift=1em] at (0.5,1) {\tikz[x=1em,y=1em,baseline=0.2em]{ \draw [fill=Green,fill opacity=0.2,Green,thick] (0,0) rectangle (1,1); } $= \int h < \int c =$ \tikz[x=1em,y=1em,baseline=0.2em]{ \draw [fill=Red,fill opacity=0.2,Red,thick] (0,0) rectangle (1,1); }};

                \node [yshift=1cm] at (0.5,1) {\textsc{Perceived participation}};
            \end{scope}

            \begin{scope}[xshift=10cm]
                \begin{scope}[yshift=2.2cm,blockchain,x=2em,y=-2em]
                    \node [block,minimum size=1.2em] (G) at (0,0) {$B_0$};
                    \foreach \i in {1,...,3} {
                        \node [block-hon] (H\i) at (-1,\i) {};
                    }
                    \foreach \i in {2,3} {
                        \pgfmathtruncatemacro{\prev}{\i-1}
                        \draw [link] (H\i) -- (H\prev);
                    }
                    \draw [link] (H1) -- (G);
                    \foreach \i in {1,...,4} {
                        \node [block-adv] (A\i) at (1,\i) {};
                    }
                    \foreach \i in {2,3,4} {
                        \pgfmathtruncatemacro{\prev}{\i-1}
                        \draw [link] (A\i) -- (A\prev);
                    }
                    \draw [link] (A1) -- (G);
                    \draw [|<->|] ([xshift=-1em]H1.north west) -- ([xshift=-1em]H3.south west) node [midway,left] {$\int h$};
                    \draw [|<->|] ([xshift=1em]A1.north east) -- ([xshift=1em]A4.south east) node [midway,right] {$\int c$};
                \end{scope}

                \node [yshift=1cm] at (0,1) {\textsc{Blockchain}};
            \end{scope}

        \end{tikzpicture}
        \caption{Forward simulation attack}
        \label{fig:forward_simulation}
    \end{subfigure}
    \caption{%
        Examples of backward and forward simulation attacks. 
        (\subref{fig:forward_simulation}) Consider the presented \emph{decaying} participation pattern amongst nodes. Let $c_0=c(0)$, and let $t$ be a timeslot such that $\int\limits_{0}^t c_0>\int\limits_0^t h(t)$. Assume that corrupt players awake at time $0$ can predict the leader order up to timeslot $t$. Then the corrupt players awake at time $0$ can costlessly simulate a valid execution of the protocol with chain length $\int\limits_0^t c_0$. As $\int\limits_{0}^t c_0>\int\limits_0^t h(t)$, newly woken up honest nodes at timeslot $t+1$ will confirm blocks from the corrupt chain, thus violating safety.
        (\subref{fig:backward_simulation}) A similar attack can be carried out by an adversary under \emph{increasing} participation, but ``in reverse'', using the players awake at a time slot $t$ in which $c_0$ corrupt players are awake such that $\int\limits_{0}^t c_0>\int\limits_0^t h(t)$ holds, this time costlessly simulating a chain as if they were all awake from time $0$.
    }
    \label{fig:simulation_attacks}
\end{figure}
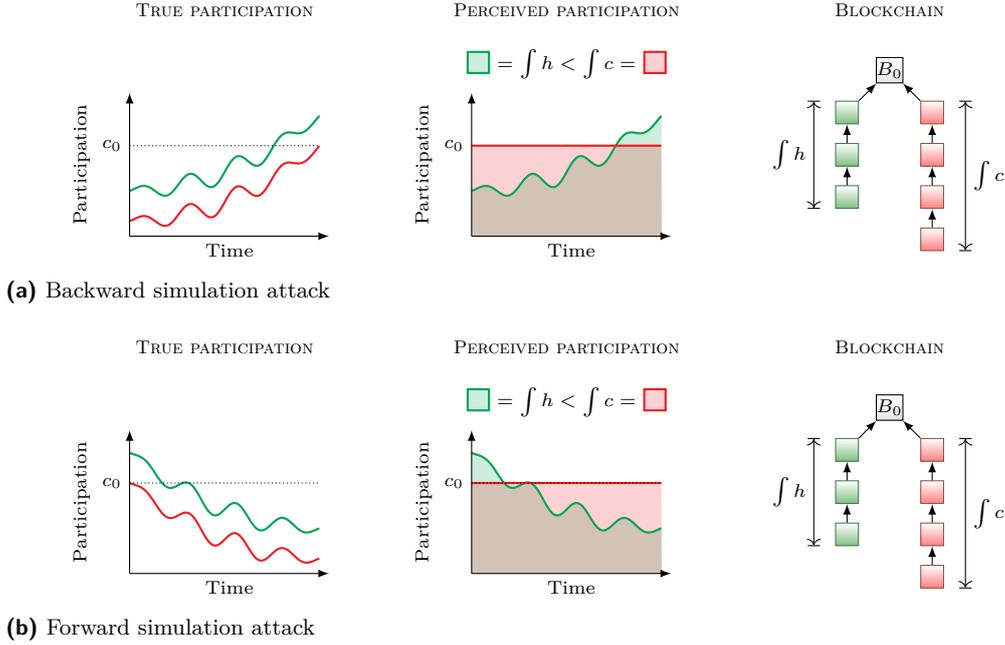

\begin{itemize}
    \item \textbf{Backward simulation.} 
    (\cref{fig:backward_simulation})
    Consider the following. The adversary commences the protocol at time $0$ with low participation of both corrupt and honest nodes. As time goes on, the adversary increases participation of corrupt (and respectively, honest) nodes until at some point in time $t$, corrupt participation at time $t$ exceeds \emph{honest} participation at timeslot $0$ by a sufficiently large amount. At this point, the corrupt nodes, via a costless simulation attack, can construct a private chain whose length exceeds the length of the honest chain constructed thus far. From the point of view of a newly woken up honest node at time $t$, the longest chain rule will cause it to build on top of the corrupt chain, violating safety. Such an attack is a concern in periods of \emph{increasing participation.}

    \item \textbf{Forward simulation.} 
    (\cref{fig:forward_simulation})
    Alternatively, consider the following scenario. Participation starts out high, and as time goes on, the adversary puts corrupt (and honest, respectively) nodes to sleep until at some time $t$, the number of awake corrupt nodes at time $0$ exceeds \emph{honest} participation at time $t$ by a sufficiently large amount. Assume further that the leader order is \emph{predictable} from time $0$. The corrupt nodes awake at time $0$ can simulate an execution from time $0$ to time $t$ with a consistent participation rate equal to the number of corrupt nodes \emph{awake at time $0$}. From the point of view of a newly woken up honest node at time $t$, the longest chain rule will cause it to build on top of the corrupt chain, violating safety. Such an attack is a concern in periods of \emph{decreasing participation.}

\end{itemize}

In this work, we devise several tools to deal with such attack vectors.

\subparagraph{Wakeness vectors.} 
We define a new primitive in the context of the sleepy model, which might be of independent interest, which we call \emph{wakeness vectors}. Intuitively, wakeness vectors allow nodes to have an estimate on the timeslots in which other nodes were awake throughout the execution a protocol. A similar idea was explored by Bar-Joseph, Keidar, and Lynch~\cite{Bar-JosephKL02} in the context of crash failures. 
Both of our protocols (for the decaying setting, and the fully-fluctuating setting) 
begin with establishing that the respective models we consider allow the nodes to implement wakeness vectors of sufficient quality. The construction of these vectors employs different techniques in each of the settings. See \cref{sec:decaying_participation} and \cref{sec:non_colluding_adversary} for an in-depth overview of the main ideas behind the construction of wakeness vectors  for the decaying and fully-fluctuating settings, respectively, along with the formal proofs.
Armed with this primitive, we then show how can be used to to secure protocol against forward/backward simulation attacks, and enhance protocols in the sleepy model to remain secure amidst decaying/fully-fluctuating participation.

\subparagraph{General compiler.} 
For our protocols (\cref{thrm:decaying_participation_protocol,thrm:main_thrm}) 
we actually prove a more general statement than discussed above. We identify key properties, which most existing protocols in the literature satisfy, that are necessary for a protocol to be compatible with our techniques, and then prove that any such protocol can be augmented to be secure amidst fully-fluctuating or decaying participation, in the presence of an \myadversary  adversary.
In the next section (\cref{sec:model_and_defs}), we pour rigor into above discussion by formally defining the models and problems we consider in this work.

\begin{table}[tbp]
    \captionsetup{singlelinecheck=off}
    \centering
    \caption[]{Overview of previous work and our contributions for atomic broadcast protocols in the synchronous sleepy setting, along with the corresponding properties/assumptions. Below, resilience refers to the fraction of corrupt nodes, $f^*$ refers to the number of \emph{actual} corruptions in a given execution, $\gamma$ refers to the participation rate of all nodes, and $\lambda$ is a security parameter; $O(\cdot)^*$ refers to $O(\cdot)$ latency in the optimistic case. The specific adversary restrictions in each of the above results are as follows. See \cref{ssec:related_work} for further details. 
    
    }
    \newcolumntype{C}{>{\footnotesize}c}
    \begin{tabular}{CCCCCCC}
    \toprule
    Paper & $\Tf$ & $\Tb$ & Resilience & VDF & Latency & Adversary restrictions \\
    \midrule
    \cite{Bitcoin,GarayKL24} & 0 & 0 & \nicefrac{1}{2} & \xmark & $O(\frac{1}{\gamma} \lambda \Delta)$ & Proof of work \\
    \midrule
    \cite{PassS17,Snowwhite,Ouroboros} & $\infty$ & $\infty$ & $\nicefrac{1}{2}$ & \xmark & $O(\frac{1}{\gamma}\lambda \Delta)$ & --- \\
    \cite{GoyalLR21} & $\infty$ & $\infty$ & $\nicefrac{1}{2}$ & \xmark & $O(\lambda \Delta)$ & --- \\
    \cite{Prism} & $\infty$ & $\infty$ & $\nicefrac{1}{2}$ & \xmark & $O(\frac{\Delta}{\gamma})^*$ & --- \\
    \cite{KhanchandaniW21} & $\infty$ & $\infty$ & $\nicefrac{1}{3}$ & \xmark & $O(n)$ & \scriptsize Nodes announce going to sleep \\
    \cite{Bar-JosephKL02} & $0$ & $0$ & $\nicefrac{1}{3}$ & \xmark & $O(f^*)$ & \scriptsize Crash failures \\
    \cite{GafniL23} & $0$ & $0$ & $\nicefrac{1}{2}$ & \xmark & $O(\Delta)$ & \scriptsize Time-stamped messages \\
    \cite{MalkhiM023,StreamlineSleepy} & $\infty$ & $O(\Delta)$ & $\nicefrac{1}{2}$ & \xmark & $O(\Delta)$ & --- \\
    \cite{MalkhiM023} & $O(\Delta)$ & $O(\Delta)$ & $\nicefrac{1}{3}$ & \xmark & $O(\Delta)$ & \scriptsize Message timeout \\
    \cite{Posat} & $O(\frac{\Delta}{\gamma})$ & $O(\frac{\Delta}{\gamma})$ & $\frac{1}{1+e}\simeq 0.268$ & \cmark & $O(\frac{1}{\gamma}\lambda \Delta)$ & \scriptsize \Myadversary adversary \\
    \cite{Goldfish} & $\infty$ & $O(\Delta)$ & $\nicefrac{1}{2}$ & \xmark & $O(\Delta)^*$ & --- \\
    \midrule
    This paper & $O(\Delta \log ^2 \lambda)$ & $\infty$ & $\nicefrac{1}{2}$ & \xmark & $O(\Delta)$ & \scriptsize \Myadversary adversary \\
    This paper & $O(\Delta)$ & $O(\Delta)$ & $\nicefrac{1}{2}$ & \cmark & $O(\Delta)$ & \scriptsize \Myadversary adversary\label{external} \\
    \bottomrule
    \end{tabular}
    \label{table:protocols}
\end{table}

\section{Preliminaries}
\label{sec:model_and_defs}

\subparagraph{Sleepy model.} 
We work in the extended sleepy model that allows fully-fluctuating participation of corrupt nodes as defined in~\cite{MalkhiM023}. Specifically, we operate in the permissioned setting, i.e., there exists a set $\cP$ of size $n$ which contains the identifiers of all the nodes, which are public and known to all nodes. Without loss of generality, we will assume that ${\cal P} = [n]$.  At each timeslot $t$, any node can exist in one of two states, \emph{awake} or \emph{asleep}. The number of awake nodes at timeslot $t$ is denoted by $0<n_t\leq n$. %
Asleep nodes at time $t$ are prohibited from sending messages or executing any code at time $t$, and all messages delivered to them at time $t$ are buffered until the next time step $t'>t$ in which they are awake.  Given a protocol $\Pi$, there are two types of nodes, \emph{honest} and \emph{corrupt}. Corrupt nodes are controlled by the \emph{adversary} (defined shortly) and can behave in any manner under the restrictions imposed on a given adversary.
Honest nodes are the nodes not controlled by the adversary, and they execute $\Pi$ at all timeslots. As for message delivery, we assume a synchronous network, with synchrony parameter $\Delta$ that is fixed and known to all nodes: for every timeslot $t$, if a node is awake at timeslot $t$, then we assume that they have received all messages sent to them prior to timeslot $t-\Delta$. We assume that the adversary has control over message delivery timing so long as it abides by the $\Delta$-synchrony assumption. As to not over encumber notation, we omit $\Delta$ from the list of inputs to protocols and algorithms discussed in this work. As mentioned earlier, asleep nodes receive all messages sent to them by timeslot $\max(t',t+\Delta)$, where $t'>t$ is the first timeslot when they are awake after timeslot $t+\Delta$.
Throughout this paper, we assume that communication is point-to-point.

\subparagraph{The environment.} 
We consider an external party to the protocol, \emph{the environment}, which has the role of providing inputs to nodes from some universe $I$ of inputs. 

\subparagraph{Oracles.}  
We adopt the notion of oracles to model ideal versions of cryptographic primitives, formally defined in~\cite{LPR23}. Such a modeling choice assists us in easily formalizing \myadversary adversary model we consider in this work, and to more cleanly state our results. Specifically, the description of a protocol also specifies a set of oracles $\cO=\set{O_1,...,O_z}$. At each timeslot $t$, each node $p_i$ may issue a finite amount of \emph{queries} to an oracle $O_j$, to which it receives a response at some timestep. An oracle's behaviour may depend on the nodes identity $i$, the current timeslot $t$, the input query $q$, and the internal randomness of the oracle, if the oracle models a random function. More formally, a deterministic oracle $O$ is modeled by a function mapping tuples $(p,m,t)\to (m,t)$
such that $O(p,m,t)=(m',t')$ implies that if $p$ queried $O$ with input $m$ at timeslot $t$, then $p$ obtained a response $m'$ at timeslot $t'$. A random oracle is modeled by a distribution over deterministic oracles.
Intuitively, in this work, oracles are going to model trusted hardware access to cryptographic primitives. From here on, we denote protocols by the tuple $(\Pi,\cO)$ where $\Pi$ is an algorithm executed by honest nodes, and $\cO$ is the set of oracles employed by $\Pi$. When $\cO$ is clear from context, we refer to a protocol as $\Pi$.

\subparagraph{Adversary and fluctuating participation.}
Throughout this paper, we focus on the \emph{static} adversary model, i.e., the identities of corrupt nodes are chosen by the adversary prior to the initiation of the protocol, and prior to any randomness being drawn, and in particular, nodes that begin the protocol as honest, cannot be corrupted later.
Denote by $f$ the total number of corrupt nodes, and by $\cF$ the set of all corrupt nodes. To formalize the notion of fully-fluctuating participation of corrupt nodes, we adopt the terminology introduced in~\cite{MalkhiM023}. Specifically, let $\cF_t$ be the  set of corrupt nodes awake at timeslot $t$, and let $f(t,\Tf,\Tb)$ be defined as follows.
\begin{gather*}
    f(t,\Tf,\Tb)=|\bigcup\limits_{t-\Tf\leq \tau \leq t+\Tb} \cF_\tau |
\end{gather*}

When $\Tf,\Tb$ are clear from context, we at times refer to $f(t,\Tf,\Tb)$ by $f_t$.
To provide some intuition for this notation, the parameters $\Tf,\Tb$ indicate how resilient the protocol has to be to forward simulation and backward simulation, respectively. In particular, for a given corrupt node $q$, if $q$ is awake at time $t$, then is it considered awake for $\Tb$ time-steps \emph{prior} to being awake, and for $\Tf$ steps \emph{after} being awake. For a given $\Tb,\Tf$, unless otherwise stated, we always assume that $f(t,\Tf,\Tb)<\frac{n}{2}$ for all $t$.

Furthermore, the adversary has complete control over the environment and over the set of nodes awake at any timeslot $t$.
While the choice of corrupt nodes occurs at the beginning of the protocol and is thus static, the same does not hold for the awake/asleep status. Namely, the set of nodes awake at each timeslot $t$ can be adaptively chosen by the adversary based on the current transcript (i.e., the contents of all messages sent by honest nodes up to, but not including, the current timeslot) of the protocol and internal state of the corrupt nodes.

\subparagraph{External adversary.} 
The view of cryptography via oracles (introduced in~\cite{LPR23}) allows us to define the \myadversary adversary model in a clean manner, in which (polynomial time) corrupt nodes do not know their secret key, and in particular, due to idealized cryptographic guarantees of the primitives implemented by the oracles, can not produce signatures on behalf of other \emph{asleep corrupt nodes}, except for with negligible probability.
This is achieved by not distributing the secret keys of nodes to them, but only allowing them oracle access to cryptographic functionalities that make use of said keys. More formally, in the \myadversary adversary model, given a protocol $(\Pi,\cO)$, each node $p$ (honest or corrupt) can only issue queries to any oracle $O\in \cO$ of the form $(p,\cdot,\cdot)$ throughout the execution of $\Pi$. In particular, corrupt nodes, even when colluding, can not obtain fresh oracle queries using the keys of \emph{asleep} corrupt nodes.

This is in contrast to the \emph{standard} adversary in which each node $p\in \cF$ can access all the secret of corrupt nodes. Rephrased in oracle terms: Every $p\in \cF$ can query any $O\in \cO$ with queries of the form $(q,\cdot,\cdot)$, for all $q\in \cF$, regardless of the asleep/awake status of $q$.

\subparagraph{Protocols and executions.} 
We are now ready to define the notion of a \emph{protocol} and an \emph{execution} of a protocol.
A protocol is specified by a pair  $(\Pi,\cO)$, where $\Pi$ is the algorithm run by the honest nodes, and $\cO$ is the set of oracles employed by $\Pi$. 
An execution of a protocol $(\Pi,\cO)$ is determined by a 4-tuple, $E=(\Pi,\cO,\cA,r)$, where $\cA$ is an adversary, and 
the vector $r$ specifies the random coins of all nodes and the adversary. An execution refers further to the contents of all messages sent and received by nodes at each timeslot. 
Given an execution $E$, we denote thus by $E^t$ the contents of the execution up to timeslot $t$, i.e., all random coins and messages received by nodes up to timeslot $t$. Given an execution $E$, a timeslot $t$, and a node $i$ awake at timeslot $t$, we denote by $E_i^t$ the \emph{view} of node $i$ of the execution $E$, which includes the internal state of $i$, the protocol $(\Pi,\cO)$, and all messages received by $i$ and the timeslots in which they were received. For a given adversary $\cA$, we define the \emph{transcript of $\Pi$ at time $t$} to be all the messages sent by honest nodes up to time $t$, and the internal state of the adversary $\cA$ at time $t$, and we denote it by $\Tr_\cA^t$.
Throughout the entire paper, we consider a finite execution horizon of $\exehorizon=\operatorname{poly}(n,\lambda)$. We assume that $\exehorizon$ is known in advance to all nodes and in particular can be taken into account in the protocol design.
We say that an execution is \emph{admissible} in the $(\Tf,\Tb,\rho)$-sleepy model iff for all timesteps $t\geq 0$ it holds that $f(t,\Tf,\Tb)<\rho n_t$.

\subparagraph{Atomic broadcast.}
\label{par:atomic_broadcast} 
We consider the atomic broadcast (AB) problem, in which the nodes are required to agree on a linearizable log of their inputs. For every timeslot $t$, every honest node $i$ reports a {\it log}, denoted by $\cL_i^t$,  of its \emph{decided values} $[x_0,x_1,x_2,...]$. %
Honest nodes make decisions about the contents of their log based on their internal state, $\Pi$, and the messages they received so far in any given execution. 
More formally, given the random variable $\bE$ over executions defined by $\Pi$ and $\cA$, $\cL_i^t(\bE^t_i)$ is the random variable that denotes the contents of the log of node $i$ at timeslot $t$, and it depends only on $\bE^t_i$. When $i$ and $t$ are clear from context, we omit them from the superscript and subscript of $\cL$.

We say that a protocol $\Pi$ solves the atomic broadcast problem with probability $1-\epsilon$ in a given execution if the following is maintained throughout the entire execution horizon.
\begin{enumerate}
    \item 
    \emph{$\epsilon$-Safety.} 
    For every timeslot $t$, and every $i,j$ such that $p_i$ and $p_j$ are honest with logs $\cL(\bE^t_i),\cL(\bE^t_j)$, with probability $1$  we have:
    $\cL(\bE_i^t)[0:m]=\cL(\bE_j^t)[0:m]$ where $m=\min\set{|\cL(\bE_i^t)|,|\cL(\bE_j^t)|}$. 

    \item 
    \emph{$(\epsilon, \ell)$-Liveness.}  
    For all $t\in [\exehorizon]$ the following holds: If input $x$ was sent to an honest node $p_i$ by the environment at timeslot $t$, then $\Pr[\forall t''\geq t+\ell\colon x\in \cL(\bE_j^{t''})]\geq 1-\epsilon$ holds for all honest nodes $p_j$.
\end{enumerate}

\subparagraph{Cryptographic primitives.} 
Let $\lambda$ be a security parameter. Throughout the paper, we make use of three cryptographic primitives. Namely, a PKI (for signatures), a VRF, and a VDF. 
We employ a 
cryptographic hash function (known and computable by all nodes), which we model as a random oracle, and denote by $\reg{H}$. 
Similarly to previous work, we denote by $\ip{m}_p$ the fact that the message $m$ was signed by node $p$. We model all three of these primitive in an \emph{idealized} fashion, using oracles $O_{\rmS}, O_{\rmV},O_{\rmD}$ that are defined as follows:
\begin{enumerate}
    \item Oracle $O_\rmS$ on a given input $(p,m,t)$ where $p$ is a node, $t$ is a timeslot, and $m$ is a message is defined to be $(\ip{m}_p, t)$. In other words, $p$ receives $\ip{m}_p$ in the same timeslot in which it issued the query. From here on in, unless explicitly said otherwise, we assume that every message being sent in any protocol by an honest node is signed by its sender.
    \item Oracle $O_\rmV$ is defined as $O_\rmV(p,m,t)=(\mathrm{VRF}_{sk_p}(m),\pi_m,t)$. In other words, $p$ receives $\mathrm{VRF}_{sk_p}(m),\pi_m$ in the same timeslot in which it issued the query. Here, $\mathrm{VRF}_{sk_p}(\cdot)$ is a VRF scheme based on $p$'s secret key $sk_p$, known by the oracle $O_\rmV$, and $\pi_m$ is the accompanying proof as per the VRF primitive. Again, to keep the focus of the paper on protocol design, we model the functionality of all the VRFs as random oracles. We further assume that the VRF functions accessible to each node all have the same domain ($\set{0,1}^*$), and the same range ($\set{0,1}^\lambda$). 

    \item Oracle $O_\rmD$ is defined as follows when $p$ sends queries $m_1,\ldots, m_k$ to the oracle:~$$O_\rmD(p,m_1,\ldots,m_k,t)=(\mathrm{VDF}(m_1),\pi_{m_1},\ldots, \mathrm{VDF}(m_k),\pi_{m_k},t).$$ In other words, $p$ receives $\mathrm{VDF}(m_i),\pi_{m_i}$ 
    at timeslot $t$. Here, $\mathrm{VDF}(m)$ is the VDF computation result on input string $m$, and $\pi_m$ is the accompanying proof as per the VDF primitive. We stipulate the following restriction on $O_\rmD$: For every round $t$ and every node $p$, $p$ can call $O_\rmD$ \emph{at most once} at timestep $t$. In other words, we capture the delay property of $O_\rmD$ by not allowing adaptive queries to $O_\rmD$ in any single timestep. This restriction applies to both honest and corrupt nodes, thus in particular, the adversary has no advantage in computing the VDF. We consider this modeling as an idealized model for a VDF.
    We model our VDF as a random function for the purposes of the analysis. We assume that the VDF accessible to each node has domain~$\set{0,1}^*$, and range~$\set{0,1}^\lambda$. 
\end{enumerate}
This would be the place to mention, that even though we model our cryptographic primitives using oracles that execute their ideal functionality, we still assume that all corrupt nodes are computationally bounded in the sense that there exists a polynomial $z(n)$ that depends on the total number of nodes 
such that the total number of queries $Q_p$ to oracles performed by any node $p$ in a single timeslot satisfies $Q_p\leq z(n)$. Finally, for an event $E$ we say that $E$ holds \emph{with high probability} (w.h.p.) if $\Pr[E]\geq 1-\negl$.

\section{Tools and definitions}

\subsection{Wakeness vectors}
\label{ssec:wakeness_vectors}

 Consider vectors $v^1,...,v^n$ whose dimension is $\exehorizon$. These vectors are defined as follows for each timeslot $t$: For all $i\in [n]$ and all $j\in [t]$, $v^i[j]=1$ if and only if node $p_i$ was awake at timeslot $j$, and $0$ otherwise. Endowing honest nodes with such vectors should intuitively make protocol design in the sleepy model substantially easier.
 Our upper bounds are constructed by having nodes implement subroutines that allow them to emulate an approximate version of this ideal functionality. The formal definition is as follows.

\begin{definition}\label{def:sound_wakeness_vectors}
    Let $(\Pi,\cO)$ be a protocol and let $\bE_{\Pi,\cA}$  be the random variable indicating a sample from the distribution over executions of $\Pi$ defined by adversary $\cA$. For a given $T$, we say that $v_p\in \set{0,1}^T$ is a \emph{$d$-valid wakeness vector} of node $p$ with respect to $\bE$ iff, except with negligible probability, $v_p[j]=1$ implies that $p$ is awake at a timeslot $j'\geq j-d$. Furthermore, if $p$ is honest, then $p$ being awake at timeslot $j$ implies $v_p[j]=1$. When $v$ is a $0$-valid wakeness vector, we simply refer to it as valid.
\end{definition}

Throughout the paper, we assume without loss of generality (w.l.o.g.) that all honest nodes keep a \emph{local view} of wakeness vectors. I.e., if $(\Pi,\cO)$ is a protocol, then in any execution $E$ of $\Pi$, each honest node $p$ stores $v_p^1,...,v_p^n$, where $v_p^j\in \set{0,1}^{\exehorizon}$ for all $j$. We say that $p$ \emph{deems} node $i$ to be awake at timeslot $t$ iff $v_p^i[t]=1$. For an interval $I$ of timeslots, we at times abuse notation and use $v_p^i[I]$ to refer to all coordinates in $I$ in $v_p^i$.
The main two upper bounds we prove in this work can be formalized as a general statement about augmenting atomic broadcast protocols with wakeness vectors.  We present the formal lemmas in the appropriate sections.

 Both our upper bounds are general compiler results, that augment protocols working in the $(\infty,\infty,\rho)$-sleepy model with wakeness vectors of sufficient quality to support fully-fluctuating participation. These augmentations however don't work for arbitrary protocols. Specifically, there is a key set of properties regarding the structure of a protocol that we must assume in order for our results to go through. It turns out, however, that the vast majority of literature on protocols for the sleepy model already posses these properties. Specifically, we define notions that we refer to as \emph{Statelessness} and \emph{Unpredictability}.

 \subparagraph{Stateless protocols.}
Intuitively, a protocol is said to be stateless if an honest nodes action depends on messages it received from nodes it has \emph{deemed} to be awake in recent timeslots. We assume w.l.o.g.\ that in any protocol $\Pi$, honest node keep a local view of wakeness vectors. Note that these vectors are not necessarily $d$-valid for any $d$. Such a property depends on the model and protocol in question. We further assume w.l.o.g.\ that honest nodes always attach the current timeslot to their messages.
\begin{restatable}{definition}{statelessdef}\label{def:stateless_protocols}
Let $(\Pi,\cO)$ be a protocol. We say that $\Pi$ is \emph{$T$-stateless} if the following holds for all honest nodes $p$ and timeslots $t$ and pair of executions $E_0,E_1$.
Let $v_p^1,...,v_p^n$ be the wakeness vectors viewed by $p$ at timeslot $t$. Let $ T'\leq T$, chosen by $p$, 
 and consider the set $S=\set{i\in [n]\mid \exists j\in [t-T',t-1], v_p^i[j]=1}$. 
 Then if $S$ is the same in both $E_0,E_1$, and furthermore, the messages sent from nodes in $S$ to $p$ are the same in both $E_0,E_1$,  then the distribution over actions of $p$ is the same at timeslot $t$ in both $E_0,E_1$.
 Furthermore, $\Pi$ solves atomic broadcast in any execution in which for all honest nodes $p$ and timeslots $t$, $S$ has a strict honest majority. 
\end{restatable}

\subparagraph{Unpredictability.} 
Intuitively, this property captures the adversary's ability to predict the contents of the log of honest nodes at some future timeslot.
\begin{restatable}{definition}{unpredictabilitydef}\label{def:log_unpredictability}
    Let $(\Pi,\cO)$ be a protocol that solves atomic broadcast. We say that $\Pi$ has \emph{$\alpha$-unpredictability} if the following holds for all $t$, and for all adversaries $\cA$. Let $\Tr^t_\cA$ be the transcript of the protocol up to timeslot $t$, 
    and let ${\cL_\cA \gets\cA(\Pi,\Tr^t_\cA,\alpha)}$ be an adversarial log, which we refer to as the adversary's \emph{guess}. Then for all honest nodes $p_i$, it holds that $\Pr[\cL({\bE}_i^t)\prec \cL_\cA\preceq\cL({\bE}_i^{t+\alpha})]=\negli{\lambda}$.
\end{restatable}

\section{Decaying participation}
\label{sec:decaying_participation}

Efficient protocols for the $\Tf=\infty$ regime (i.e., increasing participation) can be found in~\cite{Momose022,MalkhiM023,Goldfish}, with $\Tb$ being nearly optimal, namely with $\Tb=O(\Delta)$. In this section, we tackle the yet unstudied $\Tb=\infty$ case, in which participation is decaying. Namely, in the decaying participation model, corrupt nodes are allowed to go to sleep, but may never wake up after timeslot $0$. We show that that PKI, VRF, and an \myadversary adversary suffice to design protocols in this setting, with $\Tf=O(\Delta \lambda)$, where $\lambda$ is the security parameter, with error probability of $O(\frac{1}{2^\lambda})$. This result establishes the benefit of considering an \myadversary adversary, as an impossibility result by~\cite{MalkhiM023} established that atomic broadcast is impossible to solve in the $\Tb<\infty$ case against a standard adversary that can carry out the key transfer attack.

\subparagraph{Blocks.}
 We employ the notion of a \emph{block}. 
 Discussing blocks allows our black box protocols to be compatible with previous protocols in the literature that employ such a notion.
 We think of blocks as being built from the inputs sent by the environment to the nodes in the following way. 
In general, each block $B$ added to the chain is going to have the following structure:
Each block has the form $B=(x,\reg{H}(B'),v,e,\seed)$,
where $x$ is the content of the block. You should think of $x$ as all of the inputs sent by the environment to nodes that the block constructor has seen thus far that are not already decided. Here, $\reg{H}(B')$ is a hash of the previous block in the chain, with all honest nodes initializing their chain with the \emph{genesis block}, denoted by $B_0=(\bot,\bot,0)$. At times, protocols induce a partition of the timeslots into intervals which are referred to as \emph{views} and \emph{epochs}, when that is the case, $v$ indicates the \emph{view} in which the block was created, and $e$ is the \emph{epoch} in which the block was created. When the protocols we discuss do not make use of views and epochs, we simply omit these coordinates from the description of a block. Lastly $\seed$ is an ancillary string. Similarly, when we discuss protocols in which the $\seed$ entry is not used, we simply omit it from the block description. We denote by $\view(B)$ and $\epoch(B)$ the view and epoch, respectively, in which block $B$ is was (or claimed to be) created. We define the \emph{height} of a block to be its distance in the chain from the genesis block, and we denoted it by $h(B)$.

Note that any block $B$ defines a unique path from $B$ back to the genesis block, and thus defines a unique log. We say that a block $B$ extends a block $B'$ if $B=B'$ if or $B'$ is an ancestor of $B$. We say that two block $B,B'$ \emph{conflict} with each other if neither one of them extends the other.
We say that a block $B$ is valid if its ancestor block $B'$ is valid and $\epoch(B')<\epoch(B)$ or  ($\view(B')< \view(B)$ and $\epoch(B')=\epoch(B)$). If the view or epoch are omitted from the block, the above conditions hold vacuously.

\subparagraph{Permissible block.} 
On top of being valid, a protocol may impose additional constraints on blocks in order for them to be considered for confirmation. This usually takes the form a \emph{locally computable} (i.e., no communication required) predicate $\xi$, computable efficiently and locally by every node, so that $\xi(B)=1$ if $B$ is considered a permissible block, and $0$ otherwise. Messages involving impermissible blocks simply get ignored by honest nodes throughout the execution of the protocol.

\subsection{Graded proposal election}
\label{ssec:GPE}

 To obtain the result we require the observation that the protocol of~\cite{MalkhiM023} can be endowed with the unpredictability property with a slight modification which we discuss formally later in the section. This modification concerns a procedure called \emph{graded proposal election}, which we define here for completeness.

\begin{definition}\label{def:GPE}
    In \emph{graded proposal election} (GPE), each node $p$ has an input block $B_p$, and outputs a single pair $(B,g)$ of a block and a grade ($g\in \set{0,1}$) with the following properties.
    \begin{enumerate}
        \item \emph{Consistency.} If two honest nodes $p,q$ output blocks $B,B'$ respectively, so that $B\neq \bot,B'\neq \bot$, then $B=B'$.
        \item \emph{Graded Delivery.} If an honest node $p$ outputs $(B,1)$, then all honest nodes output $(B,*)$.
        \item \emph{Validity.} With probability (w.p.) at least $\frac{1}{2}$, all honest nodes output $(B,1)$ for some block $B$ that was the input of some honest node $p$.
        \item \emph{Integrity. } If an honest node $p$ outputs $(B,*)$, then either $B=B_0$, or there exists an honest node $p'$ that received the \emph{content} of $B$ from the environment.
    \end{enumerate}
\end{definition}

In their paper, Malkhi, Momose, and Ren~\cite{MalkhiM023} exhibit a protocol solving GPE in $O(\Delta)$ timesteps in the $(\infty,\infty,\frac{1}{2})$-sleepy model.

\subsection{Protocol and proof}
\label{ssec:protocol_decaying_participation}

We now have all the tools to state our results formally. The following result employs the notions of log unpredictability (\cref{def:log_unpredictability}), and  the notion of a stateless protocol (\cref{def:stateless_protocols}).
We can now formally state the following lemma, which intuitively says that assuming an external adversary, one can turn any protocol with non-trivial unpredictability and statelessness, into a protocol that can tolerate decaying participation.

\begin{lemma}\label{lemma:wakeness_vectors_augments_AC}
    Let $(\Pi,\cO)$ be a protocol solving atomic broadcast w.p.\ $1-\epsilon$ and liveness parameter $\ell$ in the $(\infty,\infty,\rho)$-sleepy model, and assume that:
    \begin{itemize}
        \item $\Pi$ has $\alpha$-unpredictability, and
        \item $\Pi$ is $\alpha$-stateless.
    \end{itemize}
        Then assuming an \myadversary adversary, $\rho<\frac{1}{2}$, there is a protocol $(\Pi',\cO)$ that solves the atomic broadcast problem in the $(O(\alpha),\infty,\rho)$-sleepy model w.p.\ $1-\epsilon-\negl$ and liveness parameter $\ell$ . Furthermore, $\Pi'$ expected latency is upper bounded by that of $\Pi'$, and incurs an additive $O(\lambda n^2)$ factor of communication per timeslot. In particular, assuming an \myadversary adversary, one can implement $O(\alpha)$-valid wakeness vectors in the $(O(\alpha),\infty,\rho)$-sleepy model.
\end{lemma}

\Cref{lemma:wakeness_vectors_augments_AC} is established via a reduction, depicted in \cref{alg:wakeness_vectors_AC}, that constructs $\Pi'$ given blackbox access to the given protocol $\Pi$ in the premise of the lemma. Instantiating $\Pi$ in the reduction of \Cref{lemma:wakeness_vectors_augments_AC} with a modification (detailed in \cref{alg:augment_MMR_unpredictable}) of the atomic broadcast protocol of~\cite[Alg.~3]{MalkhiM023}, 
we get the following result:
\begin{theorem}\label{thrm:decaying_participation_protocol}
    Assuming PKI, VRF, and an \myadversary adversary, there is a protocol $\Pi$, solving the atomic broadcast problem w.p.\ $1-\frac{1}{2^\lambda}$ in the $(O(\Delta \lambda),\infty,\frac{1}{2})$-sleepy model with $O(\Delta)$ expected latency, and liveness parameter $\ell=O(\Delta \lambda)$.
\end{theorem}

\begin{algorithm}[tbp]
    \caption{Protocol $\Pi'$ for \cref{lemma:wakeness_vectors_augments_AC}}\label{alg:wakeness_vectors_AC}
    \begin{algorithmic}[1]
        \LineComment{Each node $p$, if it is awake, executes the following at all timeslots $t$. Time is divided into epochs, each of length $\alpha$, starting from timeslot $0$. Denote the current epoch by $e(t)$. Each node initializes a local variable $\cL_p^*=B_0$. Initialize $\cL_i^0\leftarrow B_0$. Epoch $-1$ is the same as epoch $0$. Initialize $v_p^1=...=v_p^n=0^{\exehorizon}$.}

        \LineComment{Update log} \label{line:update_log}
        \For{$e=0,...,e(t)-1$}
            \State $D^*\gets$ $\#$ of nodes $q$ so that $p$ received $\ip{decide,*,e}$ 
            \For{all blocks $B$ \emph{extending a block of epoch $e-1$ in $\cL^*$}}
                \State $v_p^i[e]=1^\alpha$ for every $p_i$ that sent a block $B$ extending a block of epoch $e-1$ in $\cL_p^*$ 
                \State $D(B)\gets \#$ of nodes $q$ so that $p$ received $\ip{decide, B',e}$ for any $B'$ extending $B$
                \If{$D(B)>D^*/2$}
                    \State Add $B$ and all its ancestors to $\cL_p^*$ \label{line:adding_blocks}
                \EndIf
            \EndFor
        \EndFor

        \LineComment{Executing $\Pi$}
        \State Execute $\Pi$ as instructed at timeslot $t$, ignoring messages from nodes $p_i$ that do not satisfy $v_p^i[e(t)-1]=1^{\alpha}$ \label{line:execute_given_protocol}
        \Upon{a block $B$ is \emph{decided} according to $\Pi$} 
            \State Decide $B$ and all its ancestors.
            \State $\cL_p^*\gets $ $B$ and all its ancestors.
        \EndUpon

        \LineComment{Decision messages}
        \State Multicast $\ip{decide,B,e(t)}_p$, where $B$ is the most recently decided block \label{line:decision_messages}
    \end{algorithmic}
\end{algorithm}

\begin{algorithm}[tbp]
    \caption{Augmenting~\cite[Alg.~3]{MalkhiM023} with unpredictablility}
    \label{alg:augment_MMR_unpredictable}
    \begin{algorithmic}[1]
        \LineComment{$\GPE{v}$ invocation}
        \If{$r\in [0,4\Delta]$}
            \State $B\leftarrow (x,H(\candidate),v(t),\textcolor{red}{e(t)},\textcolor{red}{O_\rmV(\candidate,t)})$ for some $x\in I_{p}$
            \State Execute $\GPE{v} $ with input $B$. A block $B$ is \emph{permissible} if it extends $\lock$ and has $\view(B)=v(t)$, \textcolor{red}{$\epoch(B)=e(t)$, and $O_\rmV(\candidate,t)$ consists of a valid output of $O_\rmV$ for input of the form $(p,\cdot,\cdot)$.}
        \EndIf
    \end{algorithmic}
\end{algorithm}

We next provide an overview of the main ideas that go into the proof of \cref{lemma:wakeness_vectors_augments_AC}. The key observation for the decaying participation setting against an external adversary 
 is that honest nodes can \emph{always} rely on the messages \emph{claimed} to have been sent in the earlier timeslots,
due to the $\Tb=\infty$ assumption. The rough idea then is two-fold.
\begin{enumerate}
    \item 
    \textbf{Random log.} The main challenge in the decaying participation setting is dealing with forward simulation attacks, similarly to \cref{fig:forward_simulation}.
    Our idea for handling such 
    attacks is by introducing \emph{nested} randomness to the log decided by honest nodes. Specifically, along with a proposal for a block, the proposal must include a VRF output, which is computed on the VRF output attached to the parent of the proposed block. In case the proposer is honest, this randomness is unknown to the corrupt nodes until the moment of broadcasting the proposal. As such, corrupt nodes attempting to construct a corrupt chain will not be able to do so with the randomness of honest nodes beyond the latest available block/proposal sent by honest nodes. Thus, the key is to ensure that blocks proposed by \emph{honest} nodes are decided often enough. This endows the log with an unpredictability property of the contents of the log with corrupt nodes awake in the early timeslots, but not in the current timeslot. We show that we can ensure that w.h.p., a block proposed by an honest node gets decided once every $O(\Delta\lambda)$ timeslots, hence our choice of $\Tf=O(\Delta\lambda)$.
    
    \item 
    \textbf{Tolerating forward simulation.} 
    In the $\Tf<\infty$ model, we are susceptible to forward simulation attacks, in which corrupt nodes active in early rounds send messages to honest nodes awake far in the future, pretending to have been awake all along. With the randomness introduced in the first bullet point, honest nodes joining the protocol at a late time can recover the honest log by \emph{inductively} building it, starting from the first block. The inductive argument then intuitively goes as follows.
    \begin{itemize}
        \item The correct recovery of the \emph{first} block in the log is guaranteed by the $\Tb=\infty$ assumption. Namely, the number of honest nodes awake at the commencement of the protocol is greater than the total number of corrupt nodes that are \emph{ever} going to be awake during the execution. This is the base step of the reduction.
        \item The randomness introduced to the log and the choice of $\Tf$ allows honest nodes to safely deduce the next block in the log, given correct recovery up to block $k$, for any $k$. Roughly, this follows from two observations. First, any forward simulation of the protocol constructing a corrupt log up to $O(\Delta \lambda)$ timeslots forward is going to get outvoted by honest nodes, due to the $\Tf=O(\Delta \lambda)$ assumption. Any attempt by corrupt nodes to forward simulate the protocol for $\omega(\Delta \lambda)$ timeslots is foiled by the fact that any adversarial log is almost certainly\ 
         different than the honest log, as its suffix contains no blocks proposed by honest nodes, and thus the randomness attached to it is unpredictable to corrupt nodes. Hence, honest nodes awake during the following $O(\Delta \lambda)$ timeslots are still going to outvote the fake log, due to the $\Tf=O(\Delta \lambda)$ assumption. This allows honest nodes to safely deduce block $k+1$ of the log by deciding on the block for which they observe a majority of votes.
    \end{itemize}
\end{enumerate}

The proof of \cref{lemma:wakeness_vectors_augments_AC} involves augmenting a general protocol with resilience to forward simulation, this augmentation is presented in \cref{alg:wakeness_vectors_AC}. Afterwards, in order to obtain \cref{thrm:decaying_participation_protocol}, we explain how to augment the protocol of~\cite[Alg.~3]{MalkhiM023} 
with randomness and thus endow it with $O(\Delta\log^2 \lambda)$-unpredictability (see \cref{def:log_unpredictability}).

    \begin{proof}[Proof of \cref{lemma:wakeness_vectors_augments_AC}]
    Let $\Pi$ be the protocol given in the premise of \cref{lemma:wakeness_vectors_augments_AC}. 
    The protocol we construct is described in \cref{alg:wakeness_vectors_AC}. The protocol is divided into epochs, each of length $\alpha$.
    Each node that wakes up, reconstructs the contents of the log inductively starting from the first epoch. The inductive step is then implemented by each honest node ignoring all messages regarding blocks of epoch $e$ that do not extend a block of epoch $e-1$ confirmed by the node. This is formalized as described in \cref{alg:wakeness_vectors_AC}. 
In \cref{alg:wakeness_vectors_AC}, the notation $v_p^i[e]$ corresponds to all coordinates of $v_p^i$ corresponding to epoch $e$.
First we prove the following property of \cref{alg:wakeness_vectors_AC}.

\begin{lemma}\label{claim:semi_valid_wakeness_vectors}
   Except for with negligible probability, it holds that for any timeslot $t$ and honest node $p$, $v_p^1,...,v_p^n$ are $3\alpha$-valid wakeness vectors in all $(\Tf=10\alpha,\infty,\rho)$-sleepy executions. Furthermore, it holds that at the end of iteration $k$ of \emph{update log} (See line \ref{line:update_log} of \cref{alg:wakeness_vectors_AC}),
   that $\cL_p^*$ contains a block of epoch $k$ of $\cL(\bE_p^t)$, and contains no blocks that were not decided by at least one honest node.
\end{lemma}

\begin{proof}
    We prove this by induction over the epochs. Denote by $t$ the current timeslot. For $e=0$, the wakeness vector statement is trivial, i.e., whenever $v_q^i[e=0]=1^\alpha$ for an honest node $q$ for some $i$, then node $i$ was awake at timeslot at least $0$. For $\cL_q^*$,  Let $B$ be a block in $\cL^*$ that extends the genesis block, this implies that $q$ observed a majority of \emph{decide} messages for $B$ amongst all decide messages it received for blocks extending the genesis block. This includes all the honest nodes awake during epoch $0$, by line \ref{line:decision_messages} of \cref{alg:wakeness_vectors_AC}. 
    By the $(10\alpha,\infty,\rho)$-sleepy model, the honest nodes awake during epoch $0$ exceed the total number of corrupt nodes that are ever going to be awake during the execution. This implies that $p$ received a \emph{decide} message for $B$ from at least one honest node $p$, i.e., $B\in \cL(\bE_p^{t'})$ for some timeslot $t'<t$, as required. The log $\cL_q*$ contains the genesis block, and thus contains a block of $\cL(\bE_p^t)$ of epoch $0$.
    For the induction step, consider the claim to be true for all epochs at most $\leq e$, we prove the claim for epoch $e+1$.
    Let $q$ be an honest node awake at timeslot $t$ of epoch $\geq e+1$ and let $i$ be a node such that $v_q^i[e]=1^\alpha$.
    This implies that $q$ received a \emph{decide} message from $i$ extending a block in $\cL_q^*$ of epoch $e-1$. Consider the event $E$ in which $p_i$ was last awake during epoch $e'\leq e-5$.
    By the induction hypothesis, up to epoch $e-1$, all wakeness vectors stored by all honest nodes are $3\alpha$ valid.
    In particular, due to the $(\Tf=10\alpha,\infty,\rho)$-sleepy assumption, and due to the $\alpha$-stateless property of $\Pi$, we get that at least up to the end of epoch $e-1$, $\Pi$ correctly solves the atomic broadcast problem, as by line \ref{line:execute_given_protocol} of \cref{alg:wakeness_vectors_AC}, 
    all honest nodes consider messages from a set that has an honest majority.
    This allows us to invoke the $\alpha$-unpredictability of $\Pi$, together with the second induction assumption, that assures us that $\cL_q^*$ contains a block in $\cL(\bE_q^t)$ of epoch $e-1$ to claim that the probability that $i$ could send to $q$ a block extending a block of epoch $e+1$ in $\cL_q^*$ is $\negli{\lambda,n}$. This invokes the \myadversary adversary assumption, meaning that a message observed to be from $i$ can be sent only by $i$ (or other corrupt node) at a timeslot in which $i$ is awake.
    Thus $E$ occurs with at most negligible probability. Applying a union bound over all corrupt nodes and over all timeslots during the execution horizon (both polynomial in $n,\lambda$) maintains this negligible probability, and proves the claim for the wakeness vectors. This guarantees the correctness of $\Pi$ for epoch $e+1$ by its $\alpha$-stateless property. This implies that $\cL$ is consistent between all honest nodes at timeslot $t$. Now we must prove that at iteration $e$, a block of epoch $e$ in $\cL(\bE_q^t)$ is added to $\cL_q^*$, and no blocks that are not decided by some honest node are added.
    
    Assume that for an honest node $q$ no block of epoch $e$ is added to $\cL_q^*$ by the end of iteration $e$. This implies that there exists an honest node $p$, awake at timeslot $t-1$, that doesn't have any blocks of epoch $e$ in $\cL(\bE_p^{t-1})$. This again holds by the induction assumption of $\cL^*$ for iteration $e-1$, the $3\alpha$-validity of the wakeness vectors, the $\alpha$-unpredictability of $\Pi$, and the $(10\alpha,\infty,\rho)$-sleepy execution. All these imply that $q$ is hearing from an honest majority, even when restricted to honest nodes awake at timeslot $t-1$. So if all honest nodes awake at timeslot $t-1$ had a block of epoch $e$ decided, consistency of $\Pi$ would imply that it is added to $\cL^*$ for $q$ at line \ref{line:adding_blocks} of \cref{alg:wakeness_vectors_AC}. 

    This event can only happen with negligible probability, as otherwise we contradict the $\alpha$-unpredictability of $\Pi$, as the adversary at round $t-\alpha-1$ had already seen all blocks of epoch at most $e-1$ communicated between honest nodes, including the transcript of the protocol, and all messages received by $p$, and could thus compute $\cL(\bE_p^{t-1})$ with non negligible confidence. Thus except for with negligible probability, all honest nodes awake at timeslot $t-1$ have a block of epoch $e$ decided, and thus this block is added to $\cL^*$ by $q$ at timeslot $t$ by the behaviour of the protocol and the induction assumption. Lastly, again since $q$ hears from an honest majority, we get that every block $B$ added to $\cL_q^*$ was decided by at least one honest node, i.e., added to $\cL(\bE_p^{t'})$ for some honest node $p$ at a timeslot $t'<t$. 
\end{proof}
 The $3\alpha$-validity of the wakeness vectors given by \cref{claim:semi_valid_wakeness_vectors}, combined with the $(10\alpha,\infty,\rho)$-sleepy executions, and the $\alpha$-statelessness of $\Pi$, allow us to deduce that $\Pi'$ (\cref{alg:wakeness_vectors_AC}) solves the atomic broadcast correctly as required. Notice that block confirmation in $\Pi'$ coincides with block confirmation in $\Pi$, and thus they have the same latency properties. The only added communication occurs in \emph{decision messages} part of the protocol, and each message is of length $O(\lambda)$, thus the total communication blowup per timeslot is $O(\lambda n^2)$, as required.
\end{proof}
\begin{proof}[Proof of \cref{thrm:decaying_participation_protocol}]

All that is left to obtain \cref{thrm:decaying_participation_protocol}, is to observe that the protocol (specifically, Algorithm 3) of~\cite{MalkhiM023} (henceforth \cite[Alg.~3]{MalkhiM023})
is $O(\Delta)$-stateless, works in the $(\infty,\infty,\frac{1}{2})$-sleepy model (see \cref{thrm:MMR_protocol}). It can be augmented to be $O(\log^2 \lambda)$-unpredictable with an incredibly simple change to the \emph{GPE invocation} part of the protocol, which we describe in \cref{alg:augment_MMR_unpredictable}. In words, each honest node, when inputting a block $B$ to $\GPE{v}$, simply includes in its $\seed$ a VRF computation on the current timeslot and the parent block of $B$. This ensures that the contents of each block contain a component with high entropy, thus making the logs of honest nodes unpredictable.

More formally, the augmented algorithm is described in \cref{alg:augment_MMR_unpredictable}, with the red lines indicating changes from the original pseudo-code. Due to the simplicity of the change, and not to burden the write up with the technical details of \cite[Alg.~3]{MalkhiM023},
we provide below a sketch of the proof argument.
\cite[Alg.~3]{MalkhiM023} is divided into \emph{views}, each consisting of $O(\Delta)$ timeslots, during which an attempt to agree on this next block in the log is executed. This attempt is captured by the GPE procedure, which is executed once every view.
By the properties (in particular, validity) of the GPE subroutine (see \cref{def:GPE}) and the behaviour of \cite[Alg.~3]{MalkhiM023}, w.p.\ at least $\frac{1}{2}$, at the end of the GPE subroutine, \emph{all} honest nodes decide a block proposed by an honest node into their log. Conditioned on this event, the contents of $O_\rmV(\candidate,t)$ are uniformly random and \emph{unpredictable} to any adversary not awake before the beginning of the current view. In particular, the adversary $\cA$ attempting at timeslot $t$ to guess the contents of the log of \emph{any} honest node, succeeds w.p.\ at most $\frac{1}{2^\lambda}$. As different GPE executions are independent, we get that after $O( \lambda)$ views, condition 3 in \cref{def:GPE} is invoked except w.p.\ $\frac{1}{2^\lambda}$. Thus in total, the probability that an adversary %
at timeslot $t$ can correctly guess the contents of the log of \emph{any} honest node at timeslot $t+O(\lambda\Delta)$ is at most $\frac{1}{2^\lambda}$, as required.
\end{proof}

\section{Fully-fluctuating participation}
\label{sec:non_colluding_adversary}

Finally, we move on to consider fully-fluctuating participation regime, where both $\Tf,\Tb<\infty$. We show that assuming an external adversary and a VDF, one can design efficient protocols in this setting. 

\begin{theorem}\label{thrm:main_thrm}
    Assuming a PKI, a VRF, a VDF, and an \myadversary adversary, there exists a randomized protocol  that with high probability (w.h.p.)\ solves atomic broadcast  with expected latency of $O(\Delta)$, and liveness parameter $\ell=O(\Delta \log \frac{1}{\epsilon})$ in all admissible $(O(\Delta),O(\Delta), \frac{1}{2})$-sleepy executions. 
\end{theorem}

The proof of \cref{thrm:main_thrm} is driven by two lemmas.
The first lemma (\cref{corol:VRF_chain_wakeness_vector}, stated below) is to show that there is an efficient protocol that implements valid wakeness vectors, assuming PKI, VRF, and VDF oracles, and an external adversary:
\begin{lemma}\label{corol:VRF_chain_wakeness_vector}
    Assuming a PKI, a VRF, a VDF, and an \myadversary adversary, there is a protocol with $O(\lambda n^2 \log^2 n)$ communication complexity per timeslot that w.h.p.\ implements valid wakeness vectors for all timeslots $t$ over an execution horizon of length $\exehorizon$.
\end{lemma}

The second lemma (\cref{lemma:wakenness_vectors_augments_non_colluding adversary},  stated below) is a general result that gives a black-box simulation showing how to convert {\it any} stateless protocol for atomic broadcast in the $(\infty,\Tb,\rho)$-sleepy model into an atomic broadcast protocol in the stronger $(\Tb,\Tb,\rho)$-sleepy model:
\begin{lemma}\label{lemma:wakenness_vectors_augments_non_colluding adversary}
    Let $(\Pi,\cO)$ be a protocol solving atomic broadcast w.p.\ $1-\epsilon$  in the $(\infty,\Tb,\rho)$-sleepy model with liveness parameter $\ell$. Furthermore, assume that $\Pi$ is $\Tb$-stateless (\cref{def:stateless_protocols}). Then, assuming a PKI, a VRF, an \myadversary adversary, $\rho<\frac{1}{2}$, and valid wakeness vectors, there exists a protocol $(\Pi',\cO)$ solving atomic broadcast in the $(\Tb,\Tb,\rho)$-sleepy model w.p.\ $1-\epsilon$ and liveness parameter $\ell$. Furthermore, $\Pi'$ has expected latency upper bounded by that of $\Pi$ and incurs an additive $O(\lambda n^2\log^2 n)$ communication cost per timeslot.

\end{lemma}

\begin{algorithm}[tbp]
    \caption{Constructing  protocol $\Pi'$ from $\Pi$ }\label{alg:protocol_for_personal_key_access}
    \begin{algorithmic}[1]
        \LineComment{The following is executed by every honest node $p$ at every timeslot $t$ in which it is awake. Let $T=\operatorname{poly}(\lambda,n)$ be the execution horizon. Initialize $v^i_p\gets 0^T$ for all $i\in [n]$. $\cL_i^0\gets B_0$.}
        \LineComment{Wakeness vector updates}
        \State Let $(v^1_p,...,v_p^n)$ be the local view of $p$ from the last timeslot $r$ in which $p$ was awake
        \For{$r<t'\leq t$}
            \For{all $i\in [n]$}
                \If{$p$ observes a valid $t'$-depth value with prover $p_i$}
                    \State $v_p^i[t']\leftarrow 1$\Comment{$p$ deems $p_i$ awake at timeslot $t'$}
                \Else
                    \State $v_p^i[t']\leftarrow 0$\Comment{$p$ deems $p_i$ asleep at timeslot $t'$}
                \EndIf
            \EndFor
        \EndFor
        
        \LineComment{VDF chain extension}
        \State $P\leftarrow$ set of nodes $q$ so that $p$ received a valid depth-$(t-1)$ value $\val_q$ from $q$%
            \label{line:valid_nodes}
        \State $S\leftarrow$ $O(\log^2 (n\cdot t))$ uniformly random elements $(q,\val_q)$ sampled from $P$%
            \label{line:sampling_set}
        \For{all $(q,\val_q)\in S$}
            \State Compute $O_\rmD(p,\val_q,t)$ and multicast $\ip{\mathrm{VDF}(\val_q)}_p$
        \EndFor

        \LineComment{Running the protocol}
        \State Execute $\Pi$ as instructed, and in particular, decide when $\Pi$ decides, while ignoring messages from nodes $p_i$ for which $v_p^i[t-\Tb:t-1]=0$
    \end{algorithmic}
\end{algorithm}

We prove \cref{lemma:wakenness_vectors_augments_non_colluding adversary} and \cref{corol:VRF_chain_wakeness_vector} by presenting a protocol (\cref{alg:protocol_for_personal_key_access}) that both implements wakeness vectors using VDFs in the external adversary model, and uses a given protocol $\Pi$ as a black box.
 In a high level, every node at each timeslot has three responsibilities:
\begin{enumerate}
    \item Extracting decided log so far from received messages. This computation is internal, i.e. involves no communication.
    \item Extension of wakeness vectors.
    \item Running $\Pi$.
\end{enumerate}

The idea is to build, along with the blockchain of inputs of nodes, an object we refer to as \emph{ VDF chains}, from which one can extract valid wakeness vectors for honest nodes to convince other honest nodes of them being awake at certain timeslots. This allows the honest nodes 
that wake up after a long period of sleepiness to only consider messages from nodes that were awake in recent timeslots to realize the current state of the blockchain; the $\Tb$ interval in both forward simulation and backwards simulation give us the needed majority to make sure the honest nodes can perform these tasks successfully.

The proofs of wakeness allow honest nodes to determine the identities of of nodes that were awake in the previous view, and take only their messages into account in deciding on the state of the log, using the $(\Tb,\Tb,\frac{1}{2})$-sleepy assumption regarding the permissible executions.

\cref{thrm:main_thrm}  follows from \cref{corol:VRF_chain_wakeness_vector} and \cref{lemma:wakenness_vectors_augments_non_colluding adversary},  by instantiating the protocol $\Pi$ to be the protocol from~\cite{MalkhiM023}.
Specifically, \cite[Alg.~3]{MalkhiM023}
proves the following theorem, despite not stating the stateless property explicitly.

\begin{theorem}[Theorem 1, \cite{MalkhiM023}]\label{thrm:MMR_protocol}
    Assuming a PKI and a VRF, there exists a $O(\Delta)$-stateless protocol $\Pi$ solving atomic broadcast w.p.\ $1$ with expected latency $O(\Delta)$, and liveness parameter $\ell=O(\Delta \log \frac{1}{\epsilon})$, in any admissible $(\infty,O(\Delta),\frac{1}{2})$-sleepy execution.
\end{theorem}

\cref{ssec:VRFchains} is devoted to the proofs of \cref{corol:VRF_chain_wakeness_vector}, and \cref{lemma:wakenness_vectors_augments_non_colluding adversary}.

\subsection{Constructing wakeness vectors from VDFs}
\label{ssec:VRFchains}

This section is devoted to proving \cref{corol:VRF_chain_wakeness_vector,lemma:wakenness_vectors_augments_non_colluding adversary}.
\Cref{thrm:main_thrm} then follows. 
In the \myadversary adversary model, where corrupt nodes may only access their secret keys via oracles, a VDF assumption allows any honest node $p$ to prove w.h.p.\ the  following statement for any timeslot $t$, if it is true: ``node $p$ was awake at some timeslot  $t'\geq t$''.
In order to prove this statement with VDF, we need to define a few notions first.

\begin{definition}\label{def:depth_t_value}
    Let $\val$ be some value in the range of the VDF function, and let $p$ be some node. We say that $\val$ is a valid \emph{depth-$v$} value w.r.t.\ $p$ if there exists values $\val_0,...,\val_{v-1},\val_v$, where $\val_v=\val$, and  nodes $p_0,...,p_{v-1}, p_v$, where $p_v=p$, such that following holds. 
    \begin{enumerate}
        \item For all $i\in [v]$, $\mathrm{VDF}(\val_i)=\val_{i+1}$.
        \item Node $p$ has received $\ip{O_\rmD(p_i,\val_i,\cdot)}_{p_i}$ for all $i\in [v]$.
    \end{enumerate}

    We call $p_{v-1}$ the prover, and $\val$ the \emph{v-proof}.
\end{definition}

We first note that if $p$ received a valid depth-$v$ value $\val$ from $p_{v-1}$, then $p$ knows that w.h.p.\ $p_{v-1}$ sent that value at a timeslot $v'\geq v$, and in particular was awake at a timeslot $v'\geq v$. We give a formal proof of this below.

\begin{claim}\label{claim:wakeness_vectors}
    In the presence of a \myadversary adversary, if a node $p$ receives a valid depth-$t$ value $\val$ from a node $p_{t-1}$, then except with negligible probability, $p_{t-1}$ sent $\val$ at timeslot $t'\geq t$, and was \emph{awake} at a timeslot $t''\geq t-1$.
\end{claim}

\begin{proof}
    We prove this by induction on $t$. For $t=0$, the claim is trivial since a node that wasn't awake at any time $\geq 0$ can not send any messages.
    Assume that node $p$ received a valid depth-$t$ value $\val$ from $q$, for $q$ that was not awake at any timeslot $\geq t$. Denote the corresponding nodes and values by $p_1,...,p_{t-1}=q$, $\val_0,...,\val_{t-1}$. Note that in particular, $\val_{t-1}$ is a valid depth-$(t-1)$ value w.r.t.\ $p$, sent by $p_{t-2}$, and thus by the induction assumption, it was sent by $p_{t-2}$ at timeslot $t''\geq t-1$. 

    Now assume that $\val$ was sent by $p_{t-1}$ at a timeslot $r<t$, 
    this means that $p_{t-1}$ had not yet received $\val_{t-1}$ from $p_{t-2}$. Denote by $Q$ the total number of queries to the VDF oracle made by any node by timeslot $r$, that number is bounded by $r\cdot n\cdot p(n,\Pi)$, in particular, the probability that $\val_{t-1}$ appeared as the valid output of a VDF oracle at timeslot $\leq r$ is at most $\frac{r\cdot n \cdot p(n,\Pi)}{2^n}$, and thus as long as $r=o(2^n)$, w.h.p.\ $p_{t-1}$ could not have computed $O_\rmD(p_{t-1},\val_{t-1},\cdot)$ before timeslot $r$, for any $r<t$. Thus w.h.p.\ $p_{t-1}$ could have computed $O_\rmD(p_{t-1},\val_{t-1},\cdot)=\val$ only after $p_{t-2}$ sent $\val_{t-1}$, which occurred at timestep at least $t-1$,  i.e., $p_{t-1}$ could have only queried $O_\rmD(p_{t-1},\val_{t-1},\cdot)$ at timestep at least $t-1$, and received a response at timestep at least $t$ due to the behavior of $O_\rmD$, thus $p_{t-1}$ sent $\val$ at timestep at least $t$, and was thus awake at timestep at least $t-1$, as required.
\end{proof}

\begin{proof}[Proof of \cref{corol:VRF_chain_wakeness_vector}]
The full algorithm is depicted in \cref{alg:protocol_for_personal_key_access}, specifically in the \emph{wakeness vector updates} and \emph{VDF chain extension} parts. 

With \cref{claim:wakeness_vectors} in mind, the main observation is that to prove the lemma, it suffices to prove that at every timeslot, every honest node extends a valid VDF chain of at least one other \emph{honest} node. Notice that this is necessary as well, as otherwise, consider the following case involving 3 nodes $p_1,p_2,p_3$.
\begin{enumerate}
    \item Node $p_3$ is corrupt and awake at timeslot $1$, node $p_1$ is honest and awake at timeslot $2$, node $p_2$ is honest and awake at timeslot $3$.
    \item Node $p_3$ computes $\val_1=O_\rmD(p_3,\val_0,1)$, and sends it to $p_1$, but not to $p_2$.
    \item By instructions of some protocol $\Pi$, $p_1$ computes $\val_2=O_\rmD(p_1,\val_1,2)$ and broadcasts it.
\end{enumerate}

Note now that $p_2$ by timeslot $3$ did not receive $\val_1=O_\rmD(p_3,\val_0,1)$, and thus in particular, w.h.p.\ $\val_2$ is not considered a valid depth-$2$ value w.r.t.\ to $p$, even though $p_1$ was indeed awake at timeslot $2$.

The trivial solution to this would be to instruct each honest node $p$ to extend the VDF chains it received from all nodes, thus extending all honest node's chains. However, this incurs a multiplicative $n$ cost in communication. A better approach, and the one we implement in \cref{alg:protocol_for_personal_key_access}, is to let each honest node $p$ sample $O(\log^2 (n\cdot t))$ of the nodes it received valid depth-$t$ VDF computations from, and extend those chains, thus guaranteeing that except with negligible probability, at least one of the chosen nodes is honest (due to the honest majority assumption in any admissible execution) and that for all $t$, except with negligible probability all honest nodes at all timeslots $t$ observe valid chains from all honest nodes for the appropriate timeslots they were awake.

We prove this by induction on the timeslot. The base case of $t=0$ clearly holds as all honest nodes have valid wakeness vectors. Assume correctness for $t$ and consider time $t+1$, let $p$ be an honest node awake at time $t$. By the induction assumption, $p$ had valid wakeness vectors, and so $p$ set $v_p^i[t-1]=1$ for all honest nodes $i$ awake at time $t-1$. In particular, by the behavior of the protocol, this implies that $p$ observed valid depth-$(t-1)$ w.r.t.\ all honest nodes $i$ awake at time $t-1$. Furthermore, the validness of the wakeness vectors of $p$ at timeslot $t$, the set of nodes $i$ for which $v_p^i[t-1]=1$ has an honest majority. As such, 
by lines~\ref{line:valid_nodes},~\ref{line:sampling_set}
of \cref{alg:protocol_for_personal_key_access}, when $p$ chooses $O(\log^2n)$ random nodes amongst these, the probability of an honest node \emph{not} being chosen is at most $2^{\log^2 n}=\mathsf{negl}(n)$. Thus w.h.p.\ $p$ extends in round $t$ a valid depth-$(t-1)$ chain sent by an honest node, which implies that at round $t+1$, all honest nodes see a valid depth-$t$ honest chain from $p$, and thus set $p$ as being awake in their wakeness vectors, as required. This proves the completeness of the wakeness vectors at time $t+1$, i.e., that for all honest $q$ awake at time $t+1$, and all $t'<t+1$ and all honest $p$ awake at time $t'$, we have that $v_q^p[t']=1$. The soundness of the wakeness vectors at time $t+1$ is implied by \cref{claim:wakeness_vectors}. This completes the induction argument and concludes the proof.
\end{proof}
\begin{proof}[Proof of \cref{lemma:wakenness_vectors_augments_non_colluding adversary}]
Denote the given protocol in the premise of \cref{lemma:wakenness_vectors_augments_non_colluding adversary} by $\Pi$. \cref{corol:VRF_chain_wakeness_vector} proves that \emph{wakeness vector updates} and \emph{VDF chain extension} correctly implement valid wakeness vectors. Which in turns implies, due to the $(\Tb,\Tb,\rho)$-sleepy model, that $\Pi$ is executed by every honest node by only considering messages from an honest majority at every timeslot. By the $T$-stateless property of $\Pi$, we thus have that $\Pi$ solves the atomic broadcast problem correctly.  Thus $\Pi'$ inherits $\epsilon$-Safety, and $(\epsilon,\ell)$-Liveness from $\Pi$. Protocol $\Pi'$ decides blocks when $\Pi$ decides them, thus the expected latency of $\Pi'$ can not be worse than that of $\Pi$, and implementing \emph{wakeness vector updates} and \emph{VDF chain extension} requires an additive $O(\lambda n^2 \log ^2 n)$ overhead in communication at every timeslot, as required.
\end{proof}

\section{Additional related work}
\label{ssec:related_work}

The early protocols designed for the synchronous sleepy model~\cite{PassS17,Snowwhite,Ouroboros,FitziGKR18} follow the longest chain approach of Nakamoto~\cite{Bitcoin}. While inheriting its simplicity and elegance, they tend to suffer from high latency in the worst case. Furthermore, all of these protocols assume fixed and non-fluctuating participation of corrupt nodes. In this setting, the works of~\cite{Prism,GoyalLR21} are geared at breaking the $\Omega(\frac{\lambda}{\gamma})$ latency barrier. Prism~\cite{Prism} designs a protocol whose latency is independent of $\lambda$ under a certain optimistic condition, while~\cite{GoyalLR21} designs a protocol whose latency is independent of the participation rate $\gamma$ (see \cref{table:protocols} for the concrete bounds). 
Momose and Ren~\cite{Momose022} designed the first protocol to achieve near optimal latency, namely $O(\Delta)$ in expectation. This was followed up by~\cite{MalkhiM023} in which the latency was further improved, as well as modifying the protocol to work in the \emph{growing} participation setting~\cite{Goldfish}. The protocols of~\cite{Momose022,MalkhiM023} differ from the earlier protocols in the sleepy model, by emulating the PBFT~\cite{PBFT,Hotstuff} approach to consensus, as opposed to longest chain based techniques.

\subparagraph{Adversary restrictions.} 
On the road to improve latency and supporting fluctuating participation of corrupt nodes, many works have imposed restrictions on adversary capabilities. The early work of~\cite{Bar-JosephKL02} designs a protocol that can be instantiated in the fully-fluctuating sleepy model which is secure only against crash failures, with the \emph{early decision} property.
Namely, the latency of the protocol is proportional to the \emph{actual} number of crash failures in the network. The work of~\cite{GafniL23} designs a $O(\Delta)$ expected latency protocol in the sleepy model that remains secure against fully-fluctuating participation, but restricts the adversary to only send time-stamped messages, i.e., each message sent by the adversary includes the \emph{real} timeslot in which it was sent. In particular, this implies that the adversary is not allowed to equivocate (send conflicting messages to honest nodes in a given timeslot). The work of~\cite{KhanchandaniW21} considers a model where honest nodes know whether or not they will be awake or asleep in the next timeslot, and as such may use this information during the execution of the protocol (e.g., to announce their impending sleepiness to the rest of the nodes). The paper of~\cite{Posat} does not discuss fully-fluctuating participation of corrupt nodes, but their protocol, when instantiated in the external adversary model we introduce in this paper, is secure against fully-fluctuating participation, and makes use of a VDF~\cite{boneh2018verifiable}. 
An early version of~\cite{MalkhiM023} showcased a protocol that supports fully-fluctuating participation and has optimal latency. They assume authenticated channels (i.e., identity of message sender is attached to each message), and further assume that messages that do not get delivered after $\Delta$ timeslots (because the recipients are asleep) get dropped from the network and never get delivered. This essentially prevents the adversary from performing forward simulation attacks.

\subparagraph{Permissionless setting.} 
In this paper we consider the setting where the total set of nodes allowed to participate in the protocol is fixed and known to all nodes. The work of Lewis-Pye and Roughgarden~\cite{LPR23} explores a generalized version of the sleepy model
in which the set of allowed participants in the protocol changes over time (via stake), and their identities are not known in advance to nodes.

\subparagraph{PoSAT.}
It is worth noting that an existing consensus protocol is already resilient against fully-fluctuating node participation, under our external-adversary model!  
Specifically, despite the model not being discussed or defined directly in the paper, the PoSAT protocol~\cite{Posat}, which makes use of time-based cryptography (specifically, VDFs) to achieve consensus in the growing participation setting, can be proven secure in the fully-fluctuating participation setting 
 against an external adversary. Both the resilience and latency of the protocol are sub-optimal, however, with resilience $\rho=\frac{1}{1+e}\simeq 0.268 $ and latency $O(\frac{1}{\gamma} \lambda \Delta)$,
where $\gamma$ is the participation rate of nodes, i.e., $\frac{1}{\gamma}>1$, and $\lambda$ is the security parameter. Our protocol obtains optimal resilience $\frac{1}{2}$ and expected constant latency, respectively. See \cref{sec:non_colluding_adversary} for details.

\bibliographystyle{plainurl}
\bibliography{references}

\end{document}